\documentclass[10pt]{article}
\PassOptionsToPackage{hyphens}{url}
\usepackage[letterpaper,textwidth=6.151in,textheight=8.25in]{geometry}
\usepackage[leqno]{amsmath}
\usepackage{xcolor}
\usepackage[labelfont=bf,font=small]{caption}
\usepackage[colorlinks,linkcolor=blue!60!black,citecolor=blue!60!black,urlcolor=blue!60!black,hypertexnames=false]{hyperref}
\usepackage[capitalize,nameinlink]{cleveref}
\crefname{section}{section}{sections}
\crefname{subsection}{subsection}{subsections}
\Crefname{section}{Section}{Sections}
\Crefname{subsection}{Subsection}{Subsections}
\Crefname{figure}{Figure}{Figures}
\crefformat{equation}{\textup{#2(#1)#3}}
\crefrangeformat{equation}{\textup{#3(#1)#4--#5(#2)#6}}
\crefmultiformat{equation}{\textup{#2(#1)#3}}{ and \textup{#2(#1)#3}}{, \textup{#2(#1)#3}}{, and \textup{#2(#1)#3}}
\crefrangemultiformat{equation}{\textup{#3(#1)#4--#5(#2)#6}}{ and \textup{#3(#1)#4--#5(#2)#6}}{, \textup{#3(#1)#4--#5(#2)#6}}{, and \textup{#3(#1)#4--#5(#2)#6}}
\Crefformat{equation}{#2Equation~\textup{(#1)}#3}
\Crefrangeformat{equation}{Equations~\textup{#3(#1)#4--#5(#2)#6}}
\Crefmultiformat{equation}{Equations~\textup{#2(#1)#3}}{ and \textup{#2(#1)#3}}{, \textup{#2(#1)#3}}{, and \textup{#2(#1)#3}}
\crefdefaultlabelformat{#2\textup{#1}#3}
\numberwithin{equation}{section}
\numberwithin{figure}{section}
\numberwithin{table}{section}
\newcommand{\funding}[1]{\textbf{Funding:} #1}
\newcommand{\email}[1]{\href{mailto:#1}{#1}}
\newenvironment{keywords}{\par\medskip\noindent\small\textbf{Key words.}}{\par}
\newenvironment{MSCcodes}{\par\smallskip\noindent\small\textbf{MSC codes.}}{\par\medskip}
\date{}
\usepackage{amssymb}
\usepackage{mathtools}
\usepackage{bm}
\usepackage{siunitx}
\usepackage{booktabs}
\usepackage{placeins}
\usepackage{needspace}
\usepackage{graphicx}
\usepackage{tikz}
\usetikzlibrary{positioning,fit,shapes.geometric,arrows.meta}

\title{Amortized Transport-Map Likelihoods for Hierarchical Inverse Problems with Application to Wetting\thanks{\funding{This work was funded by the Deutsche Forschungsgemeinschaft (DFG, German Research Foundation), project numbers 459970814 and 459970841, and by the Athene Young Investigator Program of Technische Universit\"at Darmstadt.}}}

\author{Henning Bonart\thanks{Technische Universit\"at Darmstadt, Department of Mechanical
  Engineering, Institute for Nano- and Microfluidics, Darmstadt, Germany
  (\email{bonart@nmf.tu-darmstadt.de}, \url{https://orcid.org/0000-0002-5026-4499}).}}

\ifpdf
\hypersetup{
  pdftitle={Amortized Transport-Map Likelihoods for Hierarchical Inverse Problems with Application to Wetting},
  pdfauthor={H. Bonart}
}
\fi

\begin{document}
\maketitle

\begin{abstract}
Hierarchical Bayesian models pool measurements, but with the forward model in the likelihood every posterior evaluation costs one forward solve per cell.
To reduce this cost, we introduce a simulation-based inference framework in which a triangular transport-map likelihood, trained once per design, serves every cell.
We apply it to infer wetting properties from photographs of a stable hole in a thin liquid film.
The hole's radius depends, through the nonlinear Young--Laplace equation, on the contact angle and capillary length.
One photograph gives rank-one information about them, so we pool many in a hierarchical model whose cells, each a substrate at one volume setting, share the apparatus geometry and the noise scale.
A design fixes the number of substrates, the volume settings and the photographs per cell.
The map's posteriors agree with direct sampling of the forward model in the wetting properties at a synthetic validation design, and less closely at a larger synthetic scaling design.
At the validation design, simulation-based calibration detects no departure for any parameter.
There, the map needs several hundred times fewer forward-model evaluations than direct sampling with a gradient-free ensemble sampler for a single posterior, and about five orders of magnitude fewer over a calibration study of 200 posteriors.
On real photographs its posterior agrees with direct sampling and is consistent with the measured radii.
Trained offline, the amortized map turns new photographs into a posterior with no forward solve.
It is thus a major step toward measuring wetting properties in situ without disturbing the liquid.
\end{abstract}



\section{Introduction}\label{sec:introduction}
The contact angle and the surface tension characterize the wetting behavior of a liquid film on a substrate.
They govern whether a liquid spreads on a solid surface, coats it or dewets from it.
Measuring these wetting properties inside a running process, such as coating or inkjet printing, needs a method that leaves the liquid undisturbed.
As an example, a thin liquid film on a superhydrophobic substrate can hold a stable circular defect, and a camera photographs this hole in the liquid film from above without contact with the liquid.
The equilibrium shape of the film follows the Young--Laplace equation, and earlier work established when a hole is stable~\cite{Taylor1973-yz,Lv2018-bt}.
The radius of the hole, one scalar per photograph, depends on the contact angle of the substrate and on the liquid's capillary length, which combines its surface tension with its density and gravity.
The inverse problem of this paper is therefore to infer the contact angle and the capillary length, together with their uncertainty, from measured radii accurately and at low computational cost.
Our previous work provides the experimental data and a first Bayesian model, whose posterior was approximated by variational inference~\cite{Bisswanger2024-qb}.
A Bayesian formulation answers an inverse problem with a posterior distribution, which quantifies the uncertainty of the inferred parameters~\cite{Stuart2010-ba}.

In our case one photograph informs the inputs of the Young--Laplace model along a single direction only, so photographs at several volume settings and of several substrates must be combined.
A hierarchical model pools the photographs systematically~\cite{Gelman2013-lq}: parameters of each substrate, such as its capillary length, are drawn from a common population, and parameters such as the noise scale are shared by every photograph.
With the forward model in the likelihood, every evaluation of the posterior of such a hierarchical model costs one forward solve per cell (one substrate at one volume setting), and a chain needs many evaluations.
The cost of inference therefore grows with the cost of the forward model.
Simulation-based inference moves the forward solves out of the sampler: it learns from simulations run before inference~\cite{Cranmer2020-fs}.
Drawn from a fixed training distribution, the simulations are independent of each other and can run in parallel.
In our case, a design fixes the number of substrates, the volume settings and the number of photographs per cell.
Because a learned likelihood takes the parameters of one cell, one fit per design is the likelihood of every cell of that design.

A neural density estimator can learn such a likelihood~\cite{Papamakarios2019-snl}.
For hierarchical models, neural posterior estimation has been adapted by conditioning on the global parameters~\cite{Rodrigues2021-hn}, and amortized posterior estimation has been applied to real data~\cite{Habermann2025-ml}.
Learned likelihoods have been shared across the subjects of a hierarchical model~\cite{Fengler2021-lan} and, as a normalizing flow sampled by Hamiltonian Monte Carlo, across the supernovae of a cosmological one~\cite{Boyd2026-flowsn}, both on synthetic data.
Triangular transport maps~\cite{Marzouk2017-mt} have given simulation-based posteriors~\cite{Baptista2023-tm}, also for a model of block-copolymer self-assembly~\cite{Baptista2024-bc}.
A simulation-based likelihood has been built from triangular maps as an explicit conditional density~\cite{Rubio2023-tp}.
Transport maps have accelerated likelihood-based inference~\cite{Parno2018-tmmcmc}, also on experimental data~\cite{Muchandimath2026-tm}.
A triangular map can be parametrized to suit the available data and the complexity of the distribution, has tractable derivatives, and evaluates a normalized density and the conditionals of that density~\cite{Ramgraber2025-tt}.
A likelihood built from such a map therefore has an analytic gradient in the parameters.

We are not aware of a transport-map likelihood learned from simulations and shared across a hierarchical model of real physical measurements.
We therefore introduce a framework for simulation-based inference in hierarchical inverse problems, in which an amortized transport-map likelihood, trained once per design, serves every cell.
We test it on synthetic datasets and demonstrate it on the Young--Laplace problem with real measurements.
Direct sampling of the forward model stays feasible in the present problem, which gives a reference for testing the map, alongside simulation-based calibration.
We find that the map's posteriors agree with direct sampling on real photographs and, in the wetting properties, at a synthetic validation design, and less closely at a larger synthetic scaling design.
At the validation design they pass simulation-based calibration, and the map needs several hundred times fewer forward-model evaluations than direct sampling with a gradient-free ensemble sampler for a single posterior.
Every further posterior at the same design adds no forward solve, which makes simulation-based calibration of the hierarchical model affordable.
We expect the framework to extend to problems in which direct sampling is infeasible, with non-Gaussian noise or a more expensive forward model, provided the map stays accurate as the number of cells grows.

\Cref{sec:forwardmodel} describes the measured radii from earlier laboratory experiments~\cite{Bisswanger2024-qb}, the Young--Laplace model, the observation model and the reduced-order model.
\Cref{sec:hierarchical} shows why one photograph is not enough and builds the hierarchical Bayesian model and the designs of the simulation studies.
\Cref{sec:likelihood} constructs the map's likelihood, its training on re-noised forward solves, the posterior sampling and the cost comparison.
\Cref{sec:agreement,sec:calibration} test the map in simulation studies, against direct sampling and by simulation-based calibration, and \cref{sec:realdata} applies it to the measured radii.

\needspace{4\baselineskip}
\section{Forward model}\label{sec:forwardmodel}

\subsection{The measured data}\label{sec:experiment}
We reuse the photographs of the static film from our previous experiments and the radii reported there~\cite{Bisswanger2024-qb}.
A superhydrophobic substrate is clamped in a circular basin whose wall ends in a sharp edge, the knife edge, which pins the rim of a TiO$_2$--water film about \qty{5}{\mm} thick.
A gas jet ruptures the film at its centre, and the film retracts to a stable circular dry patch, the hole.
A syringe pump sets the dispensed liquid volume in the basin ($V$ is that volume minus the fixed volume $V_k$ below the knife edge).
Adding liquid shrinks the hole until it collapses.
A camera above the basin photographs the film, and an adaptive intensity threshold gives the hole radius.

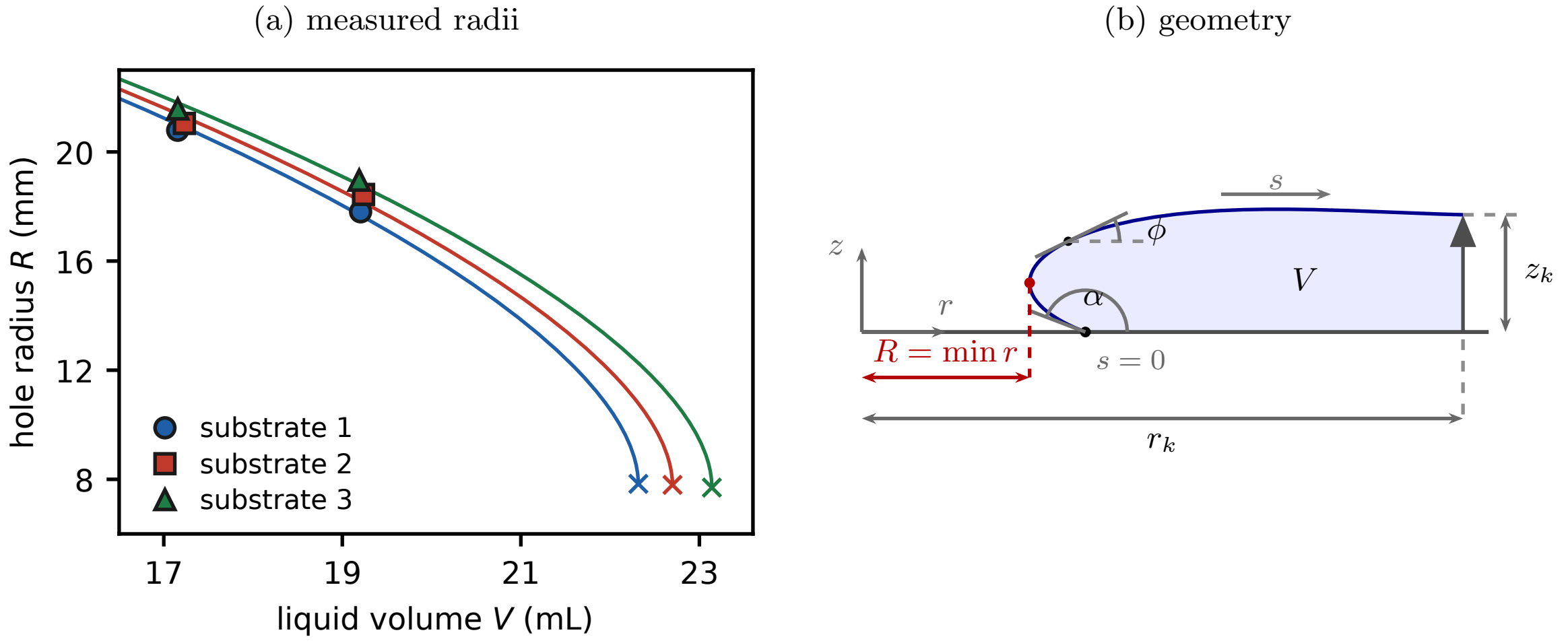
\begin{figure}[tbhp]
 \centering
 \begin{minipage}[t]{0.49\textwidth}\vspace{0pt}\centering
  {\small (a) measured radii}\par\smallskip
  \includegraphics{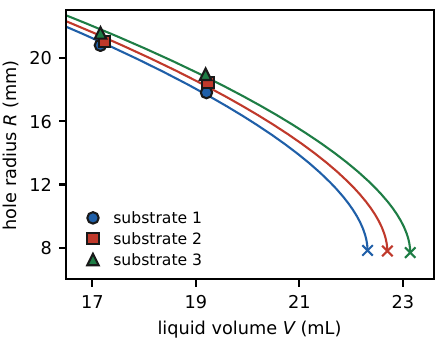}
 \end{minipage}\hfill
 \begin{minipage}[t]{0.49\textwidth}\vspace{0pt}\centering
  {\small (b) geometry}\par\vspace{12mm}
  \begin{tikzpicture}[x=0.72cm,y=0.72cm,line join=round,>={Stealth[length=4pt]}]

  \fill[blue!8]
    (3.05,0) .. controls (1.95,0.42) and (2.05,1.05) .. (3.30,1.42)
             .. controls (4.70,1.82) and (6.60,1.64) .. (8.20,1.60)
    -- (8.20,0) -- cycle;

  \draw[line width=1.1pt,black!70] (0,0) -- (8.55,0);
  \fill[black!70] (8.20,1.60) -- (8.02,1.16) -- (8.38,1.16) -- cycle;
  \draw[black!70,line width=1.0pt] (8.20,1.16) -- (8.20,0);

  \draw[line width=1.0pt,blue!55!black]
    (3.05,0) .. controls (1.95,0.42) and (2.05,1.05) .. (3.30,1.42)
             .. controls (4.70,1.82) and (6.60,1.64) .. (8.20,1.60);

  \draw[->,black!60,line width=1.0pt] (0,0) -- (1.15,0) node[above=1pt,font=\small] {$r$};
  \draw[->,black!60,line width=1.0pt] (0,0) -- (0,1.15) node[left=1pt,font=\small] {$z$};

  \fill (3.05,0) circle (1.5pt);
  \draw[black!55,line width=1.0pt] (3.05,0) -- (2.265,0.300);
  \draw[black!55,line width=1.0pt] (3.62,0) arc (0:159.1:0.57);
  \node[font=\small] at (3.16,0.46) {$\alpha$};
  \node[font=\footnotesize,black!60,anchor=north west] at (3.02,-0.08) {$s=0$};

  \draw[->,black!55,line width=1.0pt] (4.90,1.88) -- (6.40,1.88);
  \node[font=\small,black!55] at (5.65,2.04) {$s$};
  \fill (2.815,1.238) circle (1.3pt);
  \draw[dashed,black!45,line width=1.0pt] (2.815,1.238) -- (3.95,1.238);
  \draw[black!55,line width=1.0pt] (2.365,1.020) -- (3.625,1.630);
  \draw[black!55,line width=1.0pt] (3.515,1.238) arc (0:25.8:0.70);
  \node[font=\small] at (4.02,1.38) {$\phi$};

  \fill[red!70!black] (2.289,0.671) circle (1.5pt);
  \draw[red!70!black,line width=1.0pt,dashed] (2.289,0.671) -- (2.289,-0.62);
  \draw[<->,red!70!black,line width=1.0pt] (0,-0.62) -- (2.289,-0.62);
  \node[font=\small,red!70!black,anchor=south] at (1.145,-0.58) {$R=\min r$};

  \draw[dashed,black!45,line width=1.0pt] (8.20,1.60) -- (8.95,1.60);
  \draw[<->,black!60,line width=1.0pt] (8.78,0) -- (8.78,1.60);
  \node[font=\small,anchor=west] at (8.86,0.80) {$z_k$};
  \draw[<->,black!60,line width=1.0pt] (0,-1.18) -- (8.20,-1.18);
  \node[font=\small,anchor=north] at (4.10,-1.22) {$r_k$};
  \draw[dashed,black!45,line width=1.0pt] (8.20,-0.10) -- (8.20,-1.18);

  \node[font=\small] at (6.05,0.72) {$V$};

\end{tikzpicture}
 \end{minipage}
 \caption{The data and the geometry.
 (a)~Measured hole radius against the liquid volume $V$, net of the volume $V_k$ below the knife edge, for the six cells used: three substrates at two volume settings.
 Marker shape and colour identify the substrate.
 Lines, as an illustration: the full boundary-value solve for each substrate at the posterior means of direct sampling, drawn up to the largest volume, where the hole closes ($\times$).
 (b)~Schematic meridian geometry of the pinned film.
 The interface leaves the substrate at the contact angle $\alpha$, is described in arc length $s$ by its local inclination $\phi$, and is pinned at the knife edge $(r_k, z_k)$.
 The shaded region is the liquid volume $V$.
 The fixed volume $V_k$ below the edge lies outside the drawing.
  At a large contact angle the interface overhangs (exaggerated here), so the observed radius $R=\min r$ lies above the substrate.
 The layout follows Fig.~3(a) of~\cite{Bisswanger2024-qb}, the same apparatus.}
 \label{fig:experimentaldata}
\end{figure}

We use the measurements of three substrates, one per day, so substrate and day are aliased.
Each substrate enters at two volume settings.
We call one substrate at one volume setting a cell and use one photograph of the static film per cell.
\Cref{fig:experimentaldata}(a) shows the static radii analysed here.

\subsection{Young--Laplace model}\label{sec:youngla}
The equilibrium shape of the film, including the stable hole, follows the Young--Laplace equation~\cite{Lv2018-bt,Bisswanger2024-qb}.
Because the knife edge pins the film along a circle and the hole is circular, we take the interface as axisymmetric.
Its meridian profile is described by the arc length $s$ and the local inclination $\phi$ (\cref{fig:experimentaldata}(b)).
The profile balances surface tension against hydrostatic pressure,
\begin{align}
 -\sigma\left(\frac{d\phi}{ds} + \frac{\sin\phi}{r}\right) &= -\rho g z + \Delta p_0\;,\label{eqn:yl1}\\
 \frac{dr}{ds} &= \cos\phi\;,\qquad \frac{dz}{ds} = \sin\phi\;,\nonumber
\end{align}
with surface tension $\sigma$, mass density $\rho$ and gravitational acceleration $g$.
The profile leaves the substrate at the contact angle $\alpha$ and ends at the knife edge $(r_k, z_k)$: $\phi(0)=\alpha$, $z(0)=0$, $r(S)=r_k$ and $z(S)=z_k$, with $S$ the length of the profile.
The liquid volume $V$ enters through the constraint $\int_0^S 2\pi r z\,(dr/ds)\,ds = V$.
The contact-line radius $r(0)$, the length $S$ and the Laplace pressure $\Delta p_0$ at the substrate are unknown.
We solve this nonlinear boundary-value problem with a collocation solver at a given contact-line radius $r(0)$, which returns $S$, $\Delta p_0$ and the profile, and find $r(0)$ by a root search that meets the volume constraint.
Surface tension, density and gravity enter the radius only through the capillary length $l_c=\sqrt{\sigma/(\rho g)}$, so they cannot be inferred separately.
The forward model therefore takes the inputs $\bm\theta=(\alpha, l_c, V, r_k, z_k)$.

\subsection{Observation model}\label{sec:observation}
The camera sees the hole from above, and at a large (hydrophobic) contact angle the interface overhangs.
The observed radius is therefore the smallest radius of the profile, and the observation operator is $R(\bm\theta)=\min_s r(s)$.
We model a photograph as returning the radius with an additive Gaussian error,
\begin{equation}\label{eqn:observation}
 y = R(\bm\theta) + \sigma_\text{exp}\,\varepsilon\;,\qquad \varepsilon\sim\mathcal N(0,1)\;,
\end{equation}
where the noise scale $\sigma_\text{exp}$ is shared by all photographs and inferred with the other parameters.
It absorbs what the forward model does not describe: the scatter of the threshold between photographs, the variability of the procedure between runs, and the mismatch between the Young--Laplace model and the film.
The errors of the dispensed volume and of the knife-edge position enter the model as parameters of their own.

\subsection{Reduced-order model}\label{sec:rom}
Direct sampling and the training of the transport-map likelihood both call the forward model many times.
To reduce the cost of these calls, we replace the boundary-value solve by a reduced-order model.
It only has to be accurate over the inputs that inference and training visit, and we use a standard construction.
The reduced-order model expands $\phi$ in the eight leading modes of a proper orthogonal decomposition~\cite{Brunton2019-dd}, computed from full boundary-value solutions.
A least-squares solve fits their coefficients, together with $r(0)$, $S$ and $\Delta p_0$, to the residual of \cref{eqn:yl1} at interior collocation points, the boundary conditions and the volume constraint.
A radius exists only where a stable hole exists: beyond a volume that depends on the other inputs, the hole closes.
The reduced-order solve fails increasingly often as closure approaches, so we also bound the liquid volume $V$ of a cell below \qty{20}{\ml}.
The inputs at which a stable hole exists and $V$ lies below \qty{20}{\ml} form the domain of the forward model.
Off its training grid, the reduced-order model agrees with the full solve in $R$ to within about one percent up to \qty{19.2}{\ml}.
From \qty{19.6}{\ml} to \qty{21.5}{\ml}, across the domain bound and into the largest volumes of the training draws, the median difference is \qty{0.2}{\percent} and the largest about three percent (\cref{sm:rom}).

\needspace{4\baselineskip}
\section{Hierarchical Bayesian model}\label{sec:hierarchical}

\subsection{Information in the photographs}\label{sec:information}
\begin{figure}[tbhp]
 \centering
 \includegraphics[width=\textwidth]{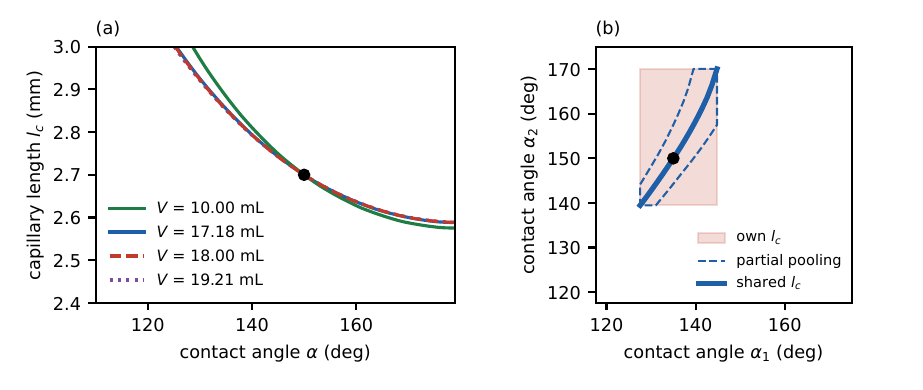}
 \caption{(a) Iso-radius curves $R(\alpha,l_c,V)=R^*(V)$ of the reduced-order model at the volumes of the validation and real-data designs, with $r_k=\qty{40}{\mm}$ and $z_k=\qty{5}{\mm}$ held.
 $R^*(V)$ is the radius at the reference point $(\qty{150}{\degree},\qty{2.7}{\mm})$ (dot), so every curve passes through it.
 The curves at the three larger volumes nearly coincide.
 (b) Contact angles of two substrates consistent with one noise-free photograph each at the smaller real-data volume, for true contact angles $(\qty{135}{\degree},\qty{150}{\degree})$ (dot) and a capillary length of \qty{2.7}{\mm}.
 Box: each substrate has a capillary length of its own, within one prior standard deviation of $\mu_{l_c}$ around \qty{2.7}{\mm}.
 Dashed: partial pooling, the two capillary lengths differ by at most \qty{0.05}{\mm}, the scale of the prior on their spread $\sigma_{l_c}$.
 Curve: the substrates share one capillary length exactly.}
 \label{fig:rankone}
\end{figure}

Consider one cell with the five shape parameters $\bm\theta=(\alpha,l_c,V,r_k,z_k)$ and one photograph $y$.
Under \cref{eqn:observation}, its Fisher information in the shape parameters is $I(\bm\theta)=\sigma_\text{exp}^{-2}\,\nabla_{\bm\theta}R\,\nabla_{\bm\theta}R^{\top}$, the outer product of the forward-model gradient with itself~\cite{Transtrum2015-sl}, which has rank one wherever $\nabla_{\bm\theta}R\neq0$.
The unknown noise scale is orthogonal to the shape parameters~\cite{Cox1987-po} and leaves this information unchanged.
A photograph therefore informs $\bm\theta$ only along $\nabla_{\bm\theta}R$ and leaves the level set $\{\bm\theta: R(\bm\theta)=y\}$ uninformed.
With $V$, $r_k$ and $z_k$ held, that level set is a curve in the plane of contact angle and capillary length.
A photograph at a second volume gives an additional curve.
The larger the angle at which the two curves cross, the more information the second photograph adds.
Unfortunately, in \cref{fig:rankone}(a), computed with the reduced-order model, the curves at the three larger volumes of the validation and real-data designs nearly coincide.
Two photographs at \qty{17.2}{\ml} and \qty{19.2}{\ml} pin the position across the curve about 270 times more tightly than the position along it, and two at \qty{10}{\ml} and \qty{18}{\ml} about 46 times more tightly.
Both ratios compare standard deviations, with $\alpha$ in units of \qty{15}{\degree} and $l_c$ in units of \qty{0.1}{\mm}, the prior widths of $\alpha$ and of the population mean of the capillary length, and they hold even with the volume and the knife edge known.
Therefore the photographs of one substrate inform mainly one combination of contact angle and capillary length.

Because the substrates share the liquid, their capillary lengths are nearly equal, and their photographs can be pooled.
With a capillary length of its own, every substrate adds a weak direction, and the photographs determine neither a contact angle nor the difference between two of them.
Tying their capillary lengths together collapses the weak directions into a single shared one: raising the capillary length and lowering every contact angle to match leaves all radii nearly unchanged.
For three substrates at the reference point and the real-data volumes, the information has three weak directions when each substrate has its own capillary length, and one when they share it (\cref{sm:information}).
The photographs then fix the contact angles up to this one direction: once one of them is known, so are the others (\cref{fig:rankone}(b)).
Because uncertain volumes and knife-edge coordinates can only lower this information, the weak directions remain in the full model.
Pooling alone thus leaves one weak direction, and an informative prior on the capillary length fixes it.
We therefore use a Bayesian hierarchical model, which combines pooling with prior knowledge.

\subsection{Hierarchy and designs}\label{sec:hierarchy}
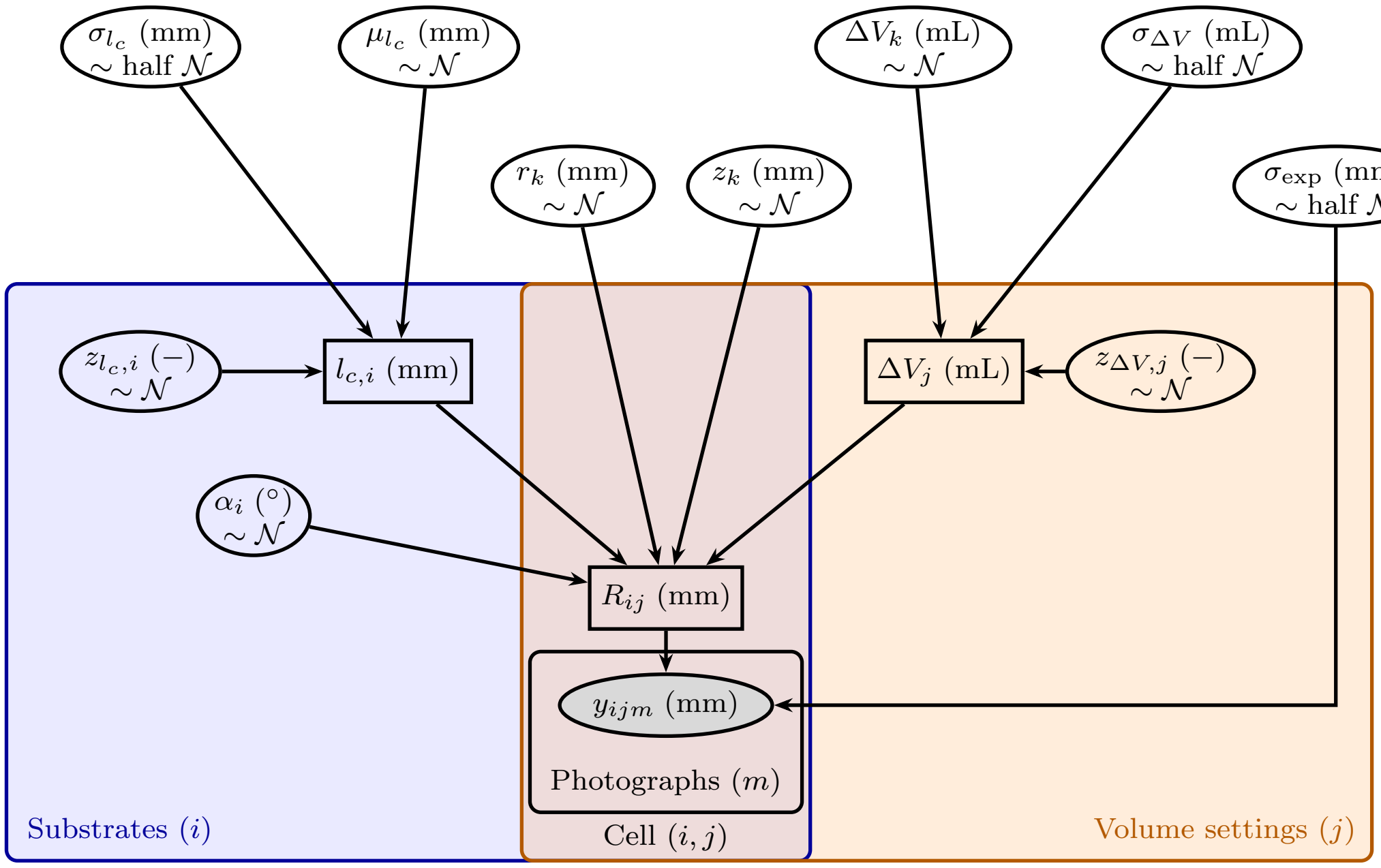
\begin{figure}[tbhp]
 \centering
 \begin{tikzpicture}[
    >={Stealth[length=1.8mm]},
    font=\footnotesize,
    line width=1pt,
    prior/.style={draw, ellipse, inner sep=0pt, minimum height=6mm, align=center},
    det/.style={draw, rectangle, inner sep=3pt, minimum height=6mm},
    obs/.style={draw, fill=gray!30, ellipse, inner sep=0.5pt, minimum height=6mm},
    plate/.style={draw, rounded corners=3pt},
    platelabel/.style={font=\footnotesize, inner sep=1pt},
    edge/.style={->},
  ]

  \fill[blue!45, fill opacity=0.18, rounded corners=3pt] (0.10, -0.15) rectangle (7.90, 5.45);
  \fill[orange!70, fill opacity=0.20, rounded corners=3pt] (5.10, -0.15) rectangle (13.35, 5.45);

  \node[prior] (siglc) at (1.50, 7.75) {$\sigma_{l_c}\;(\mathrm{mm})$\\ $\sim\text{half }\mathcal N$};
  \node[prior] (mulc)  at (4.20, 7.75) {$\mu_{l_c}\;(\mathrm{mm})$\\ $\sim\mathcal N$};
  \node[prior] (sigdv) at (11.70, 7.75) {$\sigma_{\Delta V}\;(\mathrm{mL})$\\ $\sim\text{half }\mathcal N$};
  \node[prior] (dvk)   at (8.90, 7.75) {$\Delta V_k\;(\mathrm{mL})$\\ $\sim\mathcal N$};

  \node[prior] (rk)    at (5.60, 6.40) {$r_k\;(\mathrm{mm})$\\ $\sim\mathcal N$};
  \node[prior] (zk)    at (7.50, 6.40) {$z_k\;(\mathrm{mm})$\\ $\sim\mathcal N$};
  \node[prior] (sexp)  at (13.00, 6.40) {$\sigma_\text{exp}\;(\mathrm{mm})$\\ $\sim\text{half }\mathcal N$};

  \node[prior] (zlc)   at (1.40, 4.60) {$z_{l_c,i}\;(-)$\\ $\sim\mathcal N$};
  \node[det]   (lc)    at (3.90, 4.60) {$l_{c,i}\;(\mathrm{mm})$};
  \node[prior] (alp)   at (2.50, 3.20) {$\alpha_i\;(^\circ)$\\ $\sim\mathcal N$};
  \node[det]   (dv)    at (9.20, 4.60) {$\Delta V_j\;(\mathrm{mL})$};
  \node[prior] (zdv)   at (11.30, 4.60) {$z_{\Delta V,j}\;(-)$\\ $\sim\mathcal N$};

  \node[det]   (R)     at (6.50, 2.40) {$R_{ij}\;(\mathrm{mm})$};

  \node[obs]   (y)     at (6.50, 1.36) {$y_{ijm}\;(\mathrm{mm})$};

  \draw[plate, blue!60!black] (0.10, -0.15) rectangle (7.90, 5.45);       
  \draw[plate, orange!70!black] (5.10, -0.15) rectangle (13.35, 5.45);    
  \draw[plate] (5.18, 0.32) rectangle (7.82, 1.88);                       
  \node[platelabel, anchor=south west, text=blue!60!black] at (0.25, -0.05) {Substrates $(i)$};
  \node[platelabel, anchor=south east, text=orange!70!black] at (13.25, -0.05) {Volume settings $(j)$};
  \node[platelabel, anchor=south] at (6.50, 0.42) {Photographs $(m)$};
  \node[platelabel, anchor=south] at (6.50, -0.12) {Cell $(i,j)$};

  \draw[edge] (siglc) -- (lc);
  \draw[edge] (mulc)  -- (lc);
  \draw[edge] (zlc)   -- (lc);

  \draw[edge] (sigdv) -- (dv);
  \draw[edge] (dvk)   -- (dv);
  \draw[edge] (zdv)   -- (dv);

  \draw[edge] (lc)    -- (R);
  \draw[edge] (alp)   -- (R);
  \draw[edge] (dv)    -- (R);
  \draw[edge] (rk)    -- (R);
  \draw[edge] (zk)    -- (R);

  \draw[edge] (R)     -- (y);
  \draw[edge] (sexp.south) -- (sexp.south |- 0, 1.36) -- (y.east);

\end{tikzpicture}
 \caption{Hierarchical Bayesian model as a generative model of the photographs, $y_{ijm}\sim\mathcal N(R_{ij},\sigma_\mathrm{exp}^2)$ with $R_{ij}=R(\bm\theta_{ij})$, as in \cref{eqn:likelihood}.
 The knife-edge coordinates $(r_k,z_k)$, the offset $\Delta V_k$ and the noise scale $\sigma_\mathrm{exp}$ are shared by every cell.
 The contact angle $\alpha_i$ and the capillary length $l_{c,i}$ belong to substrate $i$, and the volume error $\Delta V_j$ to volume setting $j$.
 Cell $(i,j)$ lies in both plates, where their tints overlap, and holds the radius $R_{ij}$ and the photographs $m$.
 The capillary length and the volume error are partially pooled.
 Ellipses are sampled, rectangles are deterministic given their parents, and the shaded ellipse is observed.}
 \label{fig:plate}
\end{figure}

We infer all parameters $\bm\vartheta$ jointly by Bayes' rule, $p(\bm\vartheta\mid\bm y)\propto p(\bm y\mid\bm\vartheta)\,p(\bm\vartheta)$~\cite{Gelman2013-lq}.
Substrates are indexed by $i$, volume settings by $j$ and the photographs of one cell by $m$ (\cref{fig:plate}).
Cell $(i,j)$ has the forward-model inputs $\bm\theta_{ij}=(\alpha_i, l_{c,i}, V_{0j}+\Delta V_j, r_k, z_k)$, with $V_{0j}$ the nominal volume of setting $j$, net of $V_k$ like $V$, and $\Delta V_j$ its volume error.
Under \cref{eqn:observation}, the likelihood is a product over substrates, volume settings and photographs,
\begin{equation}\label{eqn:likelihood}
 p(\bm y\mid\bm\vartheta) = \prod_i\prod_j\prod_m \mathcal N\!\left(y_{ijm}\mid R(\bm\theta_{ij}),\,\sigma_\text{exp}^2\right).
\end{equation}

Each substrate-liquid combination has its own contact angle $\alpha_i$ and capillary length $l_{c,i}$.
The capillary lengths are partially pooled through a population mean $\mu_{l_c}$ and a spread $\sigma_{l_c}$.
Each substrate was measured on its own day, and the properties of the liquid can differ slightly between days, for example through temperature or humidity.
The spread $\sigma_{l_c}$ lets the capillary lengths differ, and its prior keeps the difference small.
This partial pooling lies between a capillary length of its own for each substrate and one shared by all, the limit $\sigma_{l_c}\to0$~\cite{Gelman2013-lq} (\cref{fig:rankone}(b)).
The volume errors are partially pooled through a spread $\sigma_{\Delta V}$ about a global offset $\Delta V_k$.
A population mean of the volume error would enter the forward model only through its sum with $\Delta V_k$, and $\Delta V_k$ absorbs it.
Both pooled quantities take the non-centred parametrization,
\begin{equation}\label{eqn:noncentred}
 l_{c,i} = \mu_{l_c} + \sigma_{l_c}\,z_{l_c,i}\;,\qquad \Delta V_j = \sigma_{\Delta V}\,z_{\Delta V,j} + \Delta V_k\;,
\end{equation}
with standard normal latents $z_{l_c,i}$ and $z_{\Delta V,j}$.
This parametrization counters the funnel geometry of hierarchical posteriors~\cite{Betancourt2015-hm}.

We use three designs.
The validation design has two substrates at two volume settings, with nominal volumes of \qty{10}{\ml} and \qty{18}{\ml}, and five photographs per cell.
At this design we test the map against direct sampling, by simulation-based calibration, under skewed noise and with a smaller training set.
The scaling design has three substrates at six volume settings, equally spaced from \qty{10}{\ml} to \qty{18}{\ml}, and five photographs per cell.
With 18 cells against the validation design's four, it tests the map on a larger hierarchy.
Both designs are synthetic.
The real-data design is the layout of the measured cells of \cref{sec:experiment}, with one photograph per cell.

\begin{table}[tbhp]
 \centering
 \caption{Priors of the hierarchical model.
 Level says what each parameter is constant over.
 Normal entries give the mean and standard deviation, and the half-normal entries give scales.}
 \label{tab:priors}
 \begin{tabular}{@{}lll@{}}
  \toprule
  Parameter & Level & Prior \\
  \midrule
  $\mu_{l_c}$         & global                 & $\mathcal N(\qty{2.7}{\mm},\ \qty{0.1}{\mm})$ \\
  $\sigma_{l_c}$      & global                 & half-normal, scale \qty{0.05}{\mm} \\
  $z_{l_c,i}$         & per substrate $i$      & $\mathcal N(0, 1)$ \\
  $\alpha_i$          & per substrate $i$      & $\mathcal N(\qty{150}{\degree},\ \qty{15}{\degree})$ on $(\qty{0}{\degree},\qty{180}{\degree})$ \\
  $\sigma_{\Delta V}$ & global                 & half-normal, scale \qty{0.3}{\ml} \\
  $z_{\Delta V,j}$    & per volume setting $j$ & $\mathcal N(0, 1)$ \\
  $r_k$               & global                 & $\mathcal N(\qty{40}{\mm},\ \qty{0.333}{\mm})$ \\
  $z_k$               & global                 & $\mathcal N(\qty{5}{\mm},\ \qty{0.333}{\mm})$ \\
  $\Delta V_k$        & global                 & $\mathcal N(0,\ \qty{0.5}{\ml})$ \\
  $\sigma_\text{exp}$ & global                 & half-normal, scale \qty{0.5}{\mm} \\
  \midrule
  $V_k$               & global                 & fixed at \qty{4.835}{\ml} \\
  \bottomrule
 \end{tabular}
\end{table}

\subsection{Priors}\label{sec:priors}
\Cref{tab:priors} specifies the priors.
On one sample substrate, the advancing and receding contact angles of the coating are \qty[uncertainty-mode=separate]{163\pm5}{\degree} and \qty[uncertainty-mode=separate]{152\pm6}{\degree}~\cite[Suppl.\ Mat.]{Bisswanger2024-qb}.
Goniometry scattered and drifted as the coatings aged, while they stayed hydrophobic.
The prior on the contact angle is therefore a generous normal distribution about \qty{150}{\degree}.
The priors on the knife-edge coordinates are centred at their nominal values and read the milling tolerance of \qty{1}{\mm} as three standard deviations.
The volume $V_k$ below the knife edge (\cref{sec:experiment}) is computed from the technical drawing of the knife edge; it depends on $(r_k,z_k)$ without a closed form, and $\Delta V_k$ absorbs that uncertainty.
The prior on $\mu_{l_c}$ is informative: the measured surface tension of the dispersion, \qty{74.3}{\milli\newton\per\metre}~\cite[Suppl.\ Mat.]{Bisswanger2024-qb}, gives, with the density of the dispersion at \qty{20}{\celsius}, a capillary length of about \qty{2.75}{\mm}, inside the prior's width of \qty{0.1}{\mm} around \qty{2.7}{\mm}.
The prior on $\sigma_\text{exp}$ is weakly informative~\cite{Gelman2006-pv}.

\FloatBarrier

\section{Amortized transport-map likelihood}\label{sec:likelihood}

\subsection{Triangular maps as conditional densities}\label{sec:maps}
We replace each factor of \cref{eqn:likelihood} by a surrogate likelihood learned from forward-model simulations~\cite{Papamakarios2019-snl}.
A transport map pushes a distribution forward to a reference distribution, here the standard Gaussian $\eta$~\cite{Marzouk2017-mt}.
We order the inputs of one cell, $\bm\theta$ and $\sigma_\text{exp}$, before its datum $y$ and take the triangular (Knothe--Rosenblatt) map
\begin{equation*}
 \mathcal T(\bm\theta,\sigma_\text{exp},y) = \big(\mathcal T_{\bm\theta}(\bm\theta,\sigma_\text{exp}),\ \mathcal T_y(\bm\theta,\sigma_\text{exp},y)\big)\;.
\end{equation*}
Its last component $\mathcal T_y$, the lower block of the map~\cite{Ramgraber2025-tt}, describes the datum given the inputs.
We call it the likelihood component.
The integrated rectifier of Baptista et al.~\cite{Baptista2023-tm} makes it increase in $y$ at fixed inputs and therefore invertible in $y$.
A change of variables then gives an explicit conditional density, the surrogate likelihood
\begin{equation}\label{eqn:sbi_likelihood}
 \log\pi(y\mid\bm\theta,\sigma_\text{exp}) = \log\mathcal N\big(\mathcal T_y(\bm\theta,\sigma_\text{exp},y);\,0,1\big) + \log\partial_y\mathcal T_y(\bm\theta,\sigma_\text{exp},y)\;.
\end{equation}
If $\mathcal T$ pushes the joint distribution of inputs and datum to $\eta$, this density is the conditional of that joint~\cite{Ramgraber2025-tt}.
Fitting minimizes the Kullback--Leibler divergence between the pushforward of the training distribution and $\eta$.
This objective separates into one term per component, and each term depends on the coefficients of its own component alone~\cite{Baptista2023-tm,Ramgraber2025-tt}.
The lower block alone then gives the conditional density~\cite{Baptista2024-bc}, and we fit only $\mathcal T_y$.

\subsection{A likelihood component with an exact noise scale}\label{sec:component}
The inputs of the likelihood component include the noise scale, because every cell shares it and the posterior infers it.
The noise in \cref{eqn:observation} is Gaussian, and the component should learn this from the simulations without assuming it.
We work in variables normalized by moments of the training set: the shape parameters become $\bm t$ and the log noise scale $\tau$, with $\bm x=(\bm t,\tau)$.
The datum becomes $y_n=(y-\mu_y)/s_y$, with $\mu_y$ and $s_y$ the mean and standard deviation of the training data.
A standard likelihood component applies the integrated rectifier to a polynomial $f(\bm x,y_n)$.
It gives $f(\bm x,0)+\int_0^{y_n} g\big(\partial_{y_n} f(\bm x,y')\big)\,dy'$, with a positive rectifier $g$.
It takes $\sigma_\text{exp}$ only through $\tau$, and its slope in $y_n$ is $g(\partial_{y_n}f)$.
The Gaussian density is exact only if this slope equals $s_y/\sigma_\text{exp}$, that is, if $\partial_{y_n}f=g^{-1}(s_y/\sigma_\text{exp})$.
For SoftPlus, $g^{-1}(u)=\log(e^u-1)$ approaches $u$ for large $u$, and this target grows like $1/\sigma_\text{exp}$, exponentially in $\tau$, as $\sigma_\text{exp}\to0$.
The derivative $\partial_{y_n}f$ of a polynomial grows at most polynomially in $\tau$.
A standard component of fixed order therefore cannot be exact for every noise scale.
A standard component fitted to the training solves of our maps, with one noisy radius per solve, misplaces the noise scale at all three designs, by up to about six direct posterior standard deviations, and fails simulation-based calibration in it at both designs tested, the validation and real-data designs (\cref{sm:stdcomp}).

Our observation model scales the noise of the simulator by $\sigma_\text{exp}$ exactly.
We therefore let the noise scale enter the likelihood component in normalized units, $\ell=\log(\sigma_\text{exp}/s_y)$, as an exact coordinate with a learned correction, outside the rectifier (\cref{fig:schematic}(a)).
This follows Lueckmann et al., who choose the output form of a learned likelihood from what is known of the simulator~\cite{Lueckmann2019-en}, and the normalizing-flow model of camera noise of Abdelhamed et al., which uses an exact noise scale with a learned correction~\cite{Abdelhamed2019-nf}.
The likelihood component then becomes
\begin{equation}\label{eqn:component}
 \mathcal T_y = B(\bm x,r)\;,\qquad r = \big(y_n-a(\bm t)\big)\exp\big(-\ell+d(\bm x)\big)\;,
\end{equation}
whose shift and exponential scale take the affine form of masked autoregressive flows~\cite{Papamakarios2017-maf}.
Here $a$ is a Hermite expansion of total order three for the mean, and $d$ a correction of total order two to the width, both orders fixed before training.
The monotone component $B$ uses the SoftPlus rectifier, has total order three in $(\bm x,r)$ and is built with the software package MParT~\cite{Parno2022-mp}.
Using \cref{eqn:sbi_likelihood}, the log-likelihood is then
\begin{equation}\label{eqn:loglik}
 \log\pi(y\mid\bm\theta,\sigma_\text{exp}) = \log\mathcal N\big(B(\bm x,r);\,0,1\big) + \log\partial_r B(\bm x,r) - \ell + d(\bm x) - \log s_y\;.
\end{equation}
The correction $d$ lets the width depart from $\sigma_\text{exp}$, and $B$ lets the shape depart from the Gaussian.
Both are fitted to the simulations in training and stay fixed during inference.
Unlike the camera gain of Abdelhamed et al., which is a known setting with one learned factor per discrete gain level, our noise scale is an unknown parameter of the posterior, and the correction $d$ is a function of all inputs.
Unlike the output forms of Lueckmann et al., which are fixed in advance when the noise model is known, the Gaussian form here is only where training starts.
The Gaussian density can be recovered as a special case with $d=0$ and $B$ the identity, in which \cref{eqn:component} becomes $(y-\mu_y-s_y a(\bm t))/\sigma_\text{exp}$, exact at every noise scale.

\subsection{Training on re-noised forward solves}\label{sec:training}

\begin{figure}[tbhp]
 \centering
 \begin{tikzpicture}[
    >={Stealth[length=2.2mm]},
    font=\footnotesize,
    every path/.style={line width=1pt},
    box/.style={draw, rectangle, line width=1pt, inner sep=3pt, minimum height=8mm,
                align=center},
    lrn/.style={box, fill=gray!15},
    io/.style={draw, rounded corners=3pt, line width=1pt, inner sep=3pt,
               minimum height=8mm, align=center},
    cell/.style={draw, rectangle, line width=1pt, inner sep=2pt, minimum height=6mm,
                 minimum width=10mm},
    edge/.style={->, line width=1pt},
    note/.style={font=\scriptsize, text=black!70, align=center},
    lab/.style={font=\small\bfseries, anchor=west},
  ]

  \node[lab] at (-1.1, 2.75) {(a)};

  \node[io] (yin) at (-0.4, 0) {$y_n$};
  \node[note, below=1mm of yin] {$(y-\mu_y)/s_y$};
  \node[io] (thin) at (-0.4, 1.9) {$\bm\theta,\ \sigma_\text{exp}$};
  \node[box, minimum width=15mm] (norm) at (1.45, 1.9) {normalize};
  \draw[edge] (thin) -- (norm);

  \node[lrn, minimum width=24mm] (shift) at (3.3, 0) {shift\\[-1pt] $y_n - a(\bm t)$};
  \node[box, minimum width=28mm] (scale) at (6.6, 0)
        {scale\\[-1pt] $\times\exp\big(-\ell + d(\bm x)\big)$};
  \node[lrn, minimum width=24mm] (B) at (10.0, 0) {monotone\\[-1pt] $B(\bm x, r)$};
  \node[io] (ref) at (13.1, 0) {$\mathcal T_y$\\[-1pt] fit: ${\sim}\,\mathcal N(0,1)$};

  \draw[edge] (yin) -- (shift);
  \draw[edge] (shift) -- (scale);
  \draw[edge] (scale) -- node[above] {$r$} (B);
  \draw[edge] (B) -- (ref);

  \draw (norm.east) -- (10.0, 1.9);
  \draw[edge] (3.3, 1.9) -- node[right, pos=0.45] {$\bm t$} (shift.north);
  \draw[edge] (6.6, 1.9) -- node[right, pos=0.45] {$\ell,\ \bm x$} (scale.north);
  \draw[edge] (10.0, 1.9) -- node[right, pos=0.45] {$\bm x$} (B.north);
  \node[note, above=0.5mm of norm] {$\bm t$, $\bm x=(\bm t,\tau)$, $\ell$};

  \node[note, below=1mm of shift] (ns) {$a$: Hermite, order 3};
  \node[note, below=1mm of scale] (nc) {$\ell$ exact;\\ $d$: Hermite, order 2};
  \node[note, below=1mm of B] (nb) {$B$: MParT,\\ SoftPlus, order 3};

  \node[draw, dashed, line width=1pt, rounded corners=4pt, inner sep=1.5mm,
        fit=(shift) (scale) (B) (ns) (nc) (nb)] (comp) {};
  \node[note, anchor=north east] at ([yshift=-0.5mm]comp.south east)
        {likelihood component $\mathcal T_y$, the last component of the map};

  \begin{scope}[yshift=-4.7cm]
  \node[lab] at (-0.9, 2.3) {(b)};

  \node[io, text width=24mm] (pool) at (0.6, 0)
        {$\bm\theta^{(n)}\sim q_{\bm\theta}$\\ $N\approx2000$ rows};
  \node[box, text width=22mm] (rom) at (3.55, 0)
        {forward solve\\ $R(\bm\theta^{(n)})$, once};
  \foreach \k/\yy in {1/0.95, 2/0.32, 3/-0.32, 4/-0.95}{
    \node[draw, rectangle, line width=1pt, inner sep=1.5pt, minimum height=5mm,
          font=\scriptsize] (n\k) at (6.55, \yy) {$(\sigma^{(n,\k)},\varepsilon^{(n,\k)})$};
    \draw[edge] (rom.east) -- (n\k.west);
  }
  \node[io, text width=24mm] (set) at (9.75, 0)
        {$y^{(n,k)}=R(\bm\theta^{(n)})$\\ $+\,\sigma^{(n,k)}\varepsilon^{(n,k)}$,\\ $NK$ rows};
  \foreach \k in {1,...,4}{\draw[edge] (n\k.east) -- (set.west);}
  \node[lrn, text width=21mm] (map) at (13.0, 0) {one map: $\mathcal T_y$ of (a),\\ $\pi(y\mid\bm\theta,\sigma_\text{exp})$};

  \draw[edge] (pool) -- (rom);
  \draw[edge] (set) -- node[above, pos=0.72, font=\footnotesize] {fit} (map);
  \node[note, above=0.5mm of n1] (knote) {$K=4$ per solve};

  \node[draw, dashed, line width=1pt, rounded corners=4pt, inner sep=2.5mm,
        fit=(pool) (rom) (knote) (n4) (set)] (offline) {};
  \node[note, anchor=north west] at ([yshift=-1mm]offline.south west) {once, before sampling};

  \foreach \i/\j/\xx in {1/1/7.4, 1/2/8.8, 2/1/10.2, 2/2/11.6}{
    \node[cell] (c\i\j) at (\xx, -2.75) {$(\i,\j)$};
  }
  \node[cell] (cij) at (13.4, -2.75) {$(i,j)$};
  \node at (12.5, -2.75) {$\cdots$};
  \foreach \c in {c11,c12,c21,c22,cij}{\draw[edge] (map.south) -- (\c.north);}
  \node[note, anchor=east, text width=40mm, align=right] at (6.6, -2.75)
        {posterior sampling: the same map at every cell and step, no forward solve};
  \end{scope}
\end{tikzpicture}
 \caption{The amortized transport-map likelihood.
 (a)~The likelihood component.
 The inputs are normalized to $\bm t$, $\bm x$ and $\ell$ (\cref{sec:component}).
 The normalized datum $y_n$ is shifted by the learned mean $a(\bm t)$ and scaled by $\exp(-\ell+d(\bm x))$, in which $\ell=\log(\sigma_\text{exp}/s_y)$ is exact and $d$ is a learned correction.
 The monotone SoftPlus component $B$, built with MParT, then gives $\mathcal T_y$, and the fit drives it toward a reference standard normal.
 The polynomials $a$ and $d$ are Hermite expansions.
 (b)~Amortization.
 Each forward solve $R(\bm\theta^{(n)})$ is reused under $K=4$ draws of noise scale and noise, and the coefficients of $a$, $d$ and $B$ in (a) are fitted to the resulting training set.
 During posterior sampling the fitted likelihood component is evaluated at every cell $(i,j)$ without a forward solve.}
 \label{fig:schematic}
\end{figure}

We train one map per design of \cref{sec:hierarchy}, each on its own training draw of parameter rows $\bm\theta^{(n)}$ from a training distribution $q_{\bm\theta}$.
In the contact angle and the capillary length, $q_{\bm\theta}$ is the marginal that the prior implies for one cell.
In the knife-edge coordinates and the volume, $q_{\bm\theta}$ is broader than the prior.
The reduced-order model solves each row once, and rows for which it returns no radius are discarded.
Each solve is reused $K=4$ times under fresh noise (\cref{fig:schematic}(b)), similar to~\cite{Dax2021-dingo, Grashorn2024-tmcf}.
Each reuse draws $\sigma_\text{exp}$, independently of $\bm\theta$, from a mixture $q_\sigma$ of a half-normal and a log-uniform distribution.
It then sets $y=R(\bm\theta^{(n)})+\sigma_\text{exp}\varepsilon$ with $\varepsilon\sim\mathcal N(0,1)$.
The training distribution thus factorizes as $q_{\bm\theta}(\bm\theta)\,q_\sigma(\sigma_\text{exp})\,p(y\mid\bm\theta,\sigma_\text{exp})$, with $p$ the observation density of \cref{eqn:observation}.
Up to a constant, the likelihood component's term of the Kullback--Leibler objective is the expectation of $-\log\pi$ under this distribution~\cite{Baptista2023-tm,Ramgraber2025-tt}.
As for sequential neural likelihood~\cite{Papamakarios2019-snl}, this expectation is
\begin{equation}\label{eqn:jd}
 J_y = \mathbb E_{q_{\bm\theta}\otimes q_\sigma}\Big[\mathcal D_\text{KL}\big(p(\cdot\mid\bm\theta,\sigma_\text{exp})\,\big\|\,\pi(\cdot\mid\bm\theta,\sigma_\text{exp})\big)\Big] + c\;,
\end{equation}
with $c$ independent of the map.
The training distribution enters \cref{eqn:jd} only as a weight, which lets it differ from the prior: the volumes of each training draw can concentrate on the volume settings of its design.
The mixture $q_\sigma$ is part of the weight, with a density that is positive on all of $(0,\infty)$ and covers the support of the prior on $\sigma_\text{exp}$.
No hyperparameter of the hierarchy enters \cref{eqn:jd}, and every cell shares the observation density of \cref{eqn:observation}, so one fit is the likelihood of every cell.
By linearity of expectation, reusing each solve under $K$ noise draws keeps the target of \cref{eqn:jd} and, at a fixed number of solves, lowers the variance of its Monte Carlo estimate.

We minimize that estimate, the mean of $-\log\pi$ over the training rows, plus a small ridge penalty, with L-BFGS-B~\cite{Byrd1995-lb}.
For fixed $a$ and $d$, fitting $B$ is the standard problem for a monotone component in $(\bm x,r)$.
For the SoftPlus rectifier, Baptista, Marzouk and Zahm prove that the local minima of this problem are global when it is posed over a function space~\cite{Baptista2023-tm}.
To the best of our knowledge, the fit of finite order has no such guarantee, nor does the joint fit of $a$, $d$ and $B$.
The fitted map can therefore depend on where training starts.
Training starts at the Gaussian special case of \cref{sec:component} ($d=0$, $B$ the identity, $a$ fitted by least squares), and $d$ and $B$ learn their departures from it.
Each map is fitted once from this start.

\subsection{Posterior sampling}\label{sec:sampling}
The log posterior of the hierarchical model is the log prior plus the surrogate log-likelihood of every photograph,
\begin{equation}\label{eqn:logpost}
 \log p(\bm\vartheta\mid\bm y) = \log p(\bm\vartheta) + \sum_i\sum_j\sum_m \log\pi\big(y_{ijm}\mid\bm\theta_{ij},\sigma_\text{exp}\big) + \text{const}\;,
\end{equation}
with each term from \cref{eqn:loglik} and the inputs $\bm\theta_{ij}$ of each cell assembled through \cref{eqn:noncentred}.
The No-U-Turn sampler~\cite{Hoffman2014-nuts} draws from \cref{eqn:logpost} without a forward solve, with its analytic gradient obtained by the chain rule through the polynomials of the map, the normalization and the non-centred parametrization.
This gradient agrees closely with central finite differences, and \cref{sm:software} gives the check and the settings of the map, its training and the sampler.
The chains of the map reach a rank-normalized potential scale reduction factor $\hat R\le1.05$~\cite{Vehtari2021-rhat} for every parameter at the chain lengths we use.
As a numerical safeguard, a proposal at which \cref{eqn:loglik} is not finite is rejected.
Beyond hole closure and above the volume bound, the reduced-order model returns no radius, and the training set has no rows there.
The map's likelihood stays finite at those inputs, extrapolated from the training rows, so the surrogate posterior can place mass where direct sampling has zero density.

We count cost in forward-model evaluations, one reduced-order solve at one cell, a measure of running time where forward solves dominate it~\cite{Conrad2016-am}.
Direct sampling solves the forward model at every cell whenever it evaluates the likelihood, for every posterior.
The map pays the forward solves of its training draw once, and its posteriors need none.
We compare the two at a bulk effective sample size~\cite{Vehtari2021-rhat} of 200 for the slowest parameter of a posterior.
Wall-clock time also depends on the implementation and the hardware, and we report it as a secondary measure under the conditions of \cref{tab:sm_wallclock}.

\section{Agreement with direct sampling}\label{sec:agreement}
\Cref{sm:software,sm:convergence} give the chain lengths and the convergence of every chain of this section, and \cref{sm:tables} its values for every parameter and map.

\subsection{The reference and the measure}\label{sec:goldstandard}
Direct sampling evaluates the likelihood of \cref{eqn:likelihood} with the reduced-order model of \cref{sec:rom}, one solve per cell for each walker of an ensemble sampler~\cite{Goodman2010-gw,Foreman-Mackey2013-mh} at every step.
The same reduced-order model generates the map's training data, so direct sampling has none of the map's approximations.
The error of the reduced-order model enters both posteriors alike, and on real photographs the noise scale absorbs it.
The direct chains reach $\hat R\le1.05$ for every parameter at the chain lengths we use, as do those of the map (\cref{sec:sampling}).
Two checks with known answers, a coordinate that enters no likelihood and a direction along which the likelihood is constant, reproduce their predicted posterior standard deviations closely in the direct chains.
The cost of direct sampling per posterior is the number of steps times the number of walkers times the number of cells.

Albert et al.\ compare samplers by the Wasserstein-1 distance between the marginal of each parameter and that of a reference posterior, in parameters scaled to the unit cube~\cite{Albert2025-ws}.
We compare the map's posterior with that of direct sampling by the same distance, divided by the standard deviation of the direct marginal, $W_1/\mathrm{sd}$.
This ratio measures a difference in units of the posterior's own uncertainty: if two marginals differ only by a shift, $W_1/\mathrm{sd}$ equals that shift in direct posterior standard deviations.
As a reference level, the split-half level, we compute the same ratio between the first and second half of the direct sample and take its maximum over the parameters of a design.
This level serves as the reference for the sampling variation of direct sampling, and we read $W_1/\mathrm{sd}$ against it.

\subsection{Agreement at two synthetic designs}\label{sec:synthetic}
We first test the map on synthetic data, generated by the reduced-order model and \cref{eqn:observation} at a known ground truth.
The model of the data is then exact, and a difference between the posteriors of the map and of direct sampling beyond the sampling variation comes from the map.
At the validation design of \cref{sec:hierarchy}, the reduced-order model gives the radii at the ground truth, and \cref{eqn:observation} adds noise of scale $\sigma_\text{exp}=\qty{0.4}{\mm}$.
At the scaling design, one synthetic dataset is drawn in the same way at its own ground truth, also at \qty{0.4}{\mm}.
Direct sampling and the map condition on the same dataset at each design.
To see how far the result depends on the training draw, after the map of \cref{sec:training} was fixed, we trained three replicate maps per design, each on an independent training draw made in the same way, with as many rows with a radius.

\begin{figure}[tbhp]
 \centering
 \includegraphics[width=\textwidth]{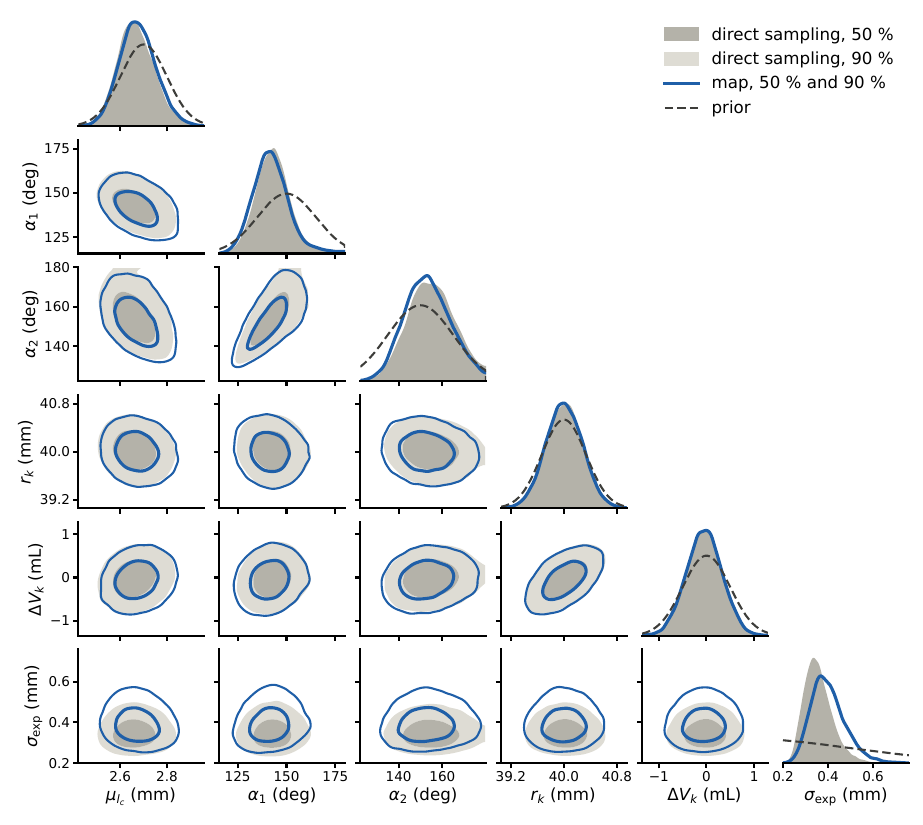}
 \caption{Pairwise joint posteriors at the validation design.
 Shaded: direct sampling, the regions holding \qty{50}{\percent} and \qty{90}{\percent} of the draws.
 Contours: the map, at the same fractions.
 Diagonal: the marginals, with direct sampling shaded, the map as a solid line and the prior as a dashed line.}
 \label{fig:triangle}
\end{figure}

\Cref{fig:triangle} shows the pairwise joint posteriors of $\mu_{l_c}$, the contact angles, $r_k$, $\Delta V_k$ and the noise scale at the validation design, with the marginals on its diagonal.
The strongest correlations lie between the two contact angles and between $r_k$ and $\Delta V_k$, with coefficients of 0.63 and 0.56 under direct sampling and 0.62 and 0.54 under the map.
The contact angles correlate negatively with $\mu_{l_c}$ ($-0.41$ and $-0.37$ under direct sampling) and positively with each other, as the weak direction of \cref{sec:information} implies.
Over all 78 pairs of parameters, the correlation coefficients differ from those of direct sampling by at most 0.14, between $\mu_{l_c}$ and the noise scale, with a median of 0.01.
\Cref{sm:tables} shows the joint posteriors at the scaling and real-data designs.

\begin{figure}[tbhp]
 \centering
 {\small (a) posterior marginals}\par
 \includegraphics[width=\textwidth]{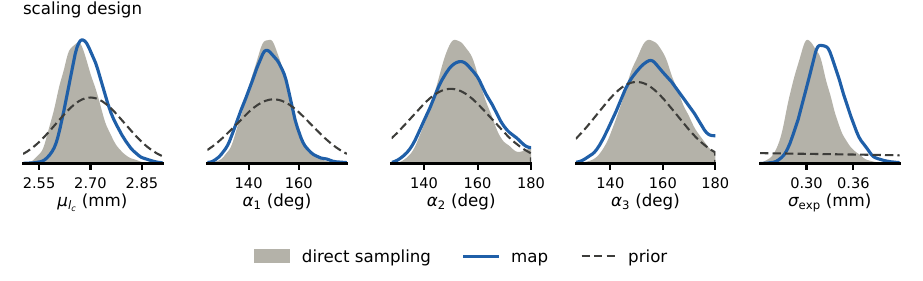}\par\medskip
 {\small (b) agreement with direct sampling}\par
 \includegraphics[width=\textwidth]{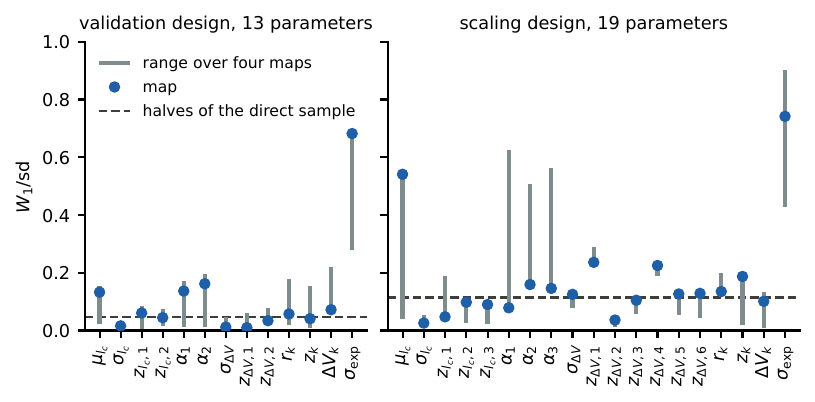}
 \caption{Agreement with direct sampling at the two synthetic designs.
 (a)~Posterior marginals of the population mean $\mu_{l_c}$ of the capillary length, the contact angles and the noise scale at the scaling design; the diagonal of \cref{fig:triangle} shows the validation design.
 \Cref{sm:tables} shows every parameter.
 Shaded: direct sampling.
 Solid line: the map.
 Dashed line: the prior.
 (b)~$W_1/\mathrm{sd}$ between the map's and direct sampling's posterior marginals, per parameter.
 Markers: the map.
 Bars: the range over the map and its three replicate maps.
 Dashed line: $W_1/\mathrm{sd}$ between the two halves of the direct sample, maximized over the parameters.
 A half holds half the draws, so this line overstates the sampling variation of the full comparison.}
 \label{fig:agreement}
\end{figure}

\Cref{fig:agreement}(a) overlays the map's marginals on those of direct sampling for the population mean $\mu_{l_c}$ of the capillary length, the contact angles and the noise scale at the scaling design.
At both designs the marginals coincide closely for every parameter except the noise scale and, at the scaling design, $\mu_{l_c}$.
The posterior of the noise scale sits higher under the map at both designs, by about two thirds of a direct posterior standard deviation at the validation design and three quarters at the scaling design.
At the scaling design, $\mu_{l_c}$ is shifted upward under the map by about half a direct posterior standard deviation ($W_1/\mathrm{sd}=0.54$).
Every other parameter lies within a few times the split-half level.

The replicate maps keep every parameter other than the noise scale within $W_1/\mathrm{sd}=0.22$ at the validation design.
At the scaling design they place the contact angles at up to 0.62, further from direct sampling than the map (\cref{fig:agreement}(b)).
In all eight maps of the two designs, the posterior mean of the noise scale lies above that of direct sampling.
Along the principal directions of the direct posterior in the contact angles and capillary lengths, the four maps differ from direct sampling by at most $W_1/\mathrm{sd}=0.21$ at the validation design.
At the scaling design all four shift the posterior mainly along one such direction, a combination of contact angles and capillary lengths, by 0.48 to 0.82 direct posterior standard deviations, and by 0.29 to 0.41 along the widest direction.
For three of the four maps the shift along the main direction exceeds $W_1/\mathrm{sd}$ of every marginal other than the noise scale.
At the scaling design, pooling the draws of the four maps gives $W_1/\mathrm{sd}$ of at most 0.45 outside the noise scale, against 0.54 for the map alone, and moves the largest marginal departure from $\mu_{l_c}$ to a contact angle.
At both synthetic designs, none of \num{1000} posterior draws of the map lies where the reduced-order model returns no radius, so this comparison is free of the extrapolation of \cref{sec:sampling}.

The map's posterior thus agrees with direct sampling at the validation design, apart from the shift of the noise scale.
On the larger hierarchy of the scaling design it agrees less closely, and every replicate map shares its shift, mainly along one combination of contact angles and capillary lengths.

\subsection{The learned noise model}\label{sec:noisemodel}
The likelihood component learns the width and the shape of the noise from the simulations (\cref{sec:component}).
To see what this costs when the noise is Gaussian, and what it gains when the noise is not, we compare it with its Gaussian special case, $d=0$ and $B$ the identity, which assumes Gaussian noise and is fitted on the same training draw.
On the Gaussian data of \cref{sec:synthetic}, the special case agrees more closely than the map in the noise scale, with $W_1/\mathrm{sd}$ of 0.17 and 0.04 at the validation and scaling designs, and in $\mu_{l_c}$ at the scaling design.
In the other parameters the two components differ from direct sampling by comparable amounts, each at most 0.3.
Along the combination of contact angles and capillary lengths that the learned maps share at the scaling design, one fit of the special case reaches $W_1/\mathrm{sd}=0.26$, against 0.48 to 0.82 (\cref{tab:sm_joint}).
At the validation design its largest value along these directions, 0.23, lies above those of the learned maps, 0.15 to 0.21.
On this Gaussian testbed, the shift of the noise scale comes mostly from learning the width and shape.
At the validation design the shift persists as the training draw grows from 250 to about \num{2000} rows.
At the map's posterior draws, the mean $\mu_y+s_ya(\bm t)$ of the likelihood component differs from the reduced-order radius by under 0.8 noise scales in every cell but one, where the reduced-order solve returns a second, much smaller root.

We repeated the validation design with skewed noise, for which the special case assumes the wrong law.
The noise follows a standardized sinh-arcsinh law~\cite{Jones2009-sinh} of skewness 1.0, so that $\sigma_\text{exp}$ remains the noise scale.
Both components are fitted to the same forward solves re-noised with that law, against direct sampling with the exact skewed likelihood.
The test covers one law, one dataset at this design and one training draw per component.
The Gaussian special case places the noise scale about one posterior standard deviation low ($W_1/\mathrm{sd}=0.94$).
The learned component agrees to 0.25, about twice the split-half level.
The special case places the noise scale near the realized standard deviation of this dataset's noise, about two thirds of the law's, so the size of its departure may depend on the dataset.

Learning the noise model thus shifts the noise scale on Gaussian data, and under the one skewed law tested it brings the noise scale closer to direct sampling than the Gaussian special case does.

\subsection{Cost of one posterior}\label{sec:cost}
We compare the map with direct sampling for a single posterior, the case least favourable to the map.
Counting cost as in \cref{sec:sampling}, we price direct chains at the validation design that stop at the effective sample size given there.
Their length is extrapolated linearly from the direct chain of \cref{sec:synthetic}.
These chains need about $10^{2.8}$ times as many forward-model evaluations as the map's training draw, or $10^{2.6}$ with their burn-in left uncounted.
At the chain lengths of direct sampling, the ratio is $10^{4.2}$ to $10^{5.0}$ over the three designs of \cref{sec:hierarchy}.
In wall-clock time, under the conditions of \cref{tab:sm_wallclock}, the map is about twice as fast as direct sampling for one posterior.
In process-hours its advantage is an order of magnitude larger.
The baseline is a gradient-free ensemble sampler, so these ratios include the difference between the two samplers as well as the forward solves the map removes.
A smaller training draw lowers the map's cost further.
At the validation design, a map trained on 250 rows agrees with direct sampling within the range spanned by the four full-budget maps of \cref{fig:agreement}(b), for every parameter.
At the real-data design, a map trained on 250 rows also agrees closely, within $W_1/\mathrm{sd}=0.12$ for every parameter.

Even for a single posterior and against this baseline, the map thus needs several hundred times fewer forward-model evaluations than direct sampling, and every further posterior at the same design adds no forward solve.

\needspace{4\baselineskip}
\section{Calibration}\label{sec:calibration}

\subsection{Simulation-based calibration}\label{sec:sbc}
We test the calibration of the surrogate posterior over many datasets drawn from the prior, by simulation-based calibration~\cite{Talts2018-vc}.
Each replication draws a ground truth from the prior, simulates a dataset by \cref{eqn:observation}, samples the posterior and records the rank of the truth among $L$ posterior draws.
Under calibration, the rank of each parameter is uniform on $\{0,\dots,L\}$.
Per parameter, we apply a chi-squared test on ten bins and a Kolmogorov--Smirnov test.
A parameter departs from uniformity when either $p$-value falls below the Bonferroni correction $0.05/n$, with $n$ the number of parameters in the study.
We also apply the ECDF test of S\"ailynoja et al.~\cite{Sailynoja2022-gu}, which compares the empirical distribution function of the ranks with simultaneous binomial bands, at the same level.
A study needs one posterior per replication, which makes it costly with direct sampling.
With one trained map for every posterior, it costs only the forward solves of the map's training draw plus one solve per cell per accepted dataset.

Our studies test the surrogate posterior as a whole, the map and the hierarchical model together.
We sample the map's posterior of \cref{eqn:logpost} and use $L=200$ posterior draws, the effective sample size of \cref{sec:sampling}.
The $L$ draws are equally spaced in the pooled draws of the four chains.
A replication enters when the smallest bulk effective sample size over its parameters reaches $L$, and replications that do not finish within a wall-clock cap are left out.
Every replication that enters also meets the $\hat R$ limit of \cref{sec:sampling}.
A ground truth is redrawn when some cell of its dataset lies outside the domain of \cref{sec:rom}, so the study covers that domain.
\Cref{sm:software} gives the caps, \cref{sm:convergence} the $\hat R$ of the replications, and \cref{sm:sbc} the statistics of every study, including the verdicts without the convergence gate.

\subsection{Results at the two designs}\label{sec:sbcresults}
\begin{figure}[tbhp]
 \centering
 \includegraphics[width=\textwidth]{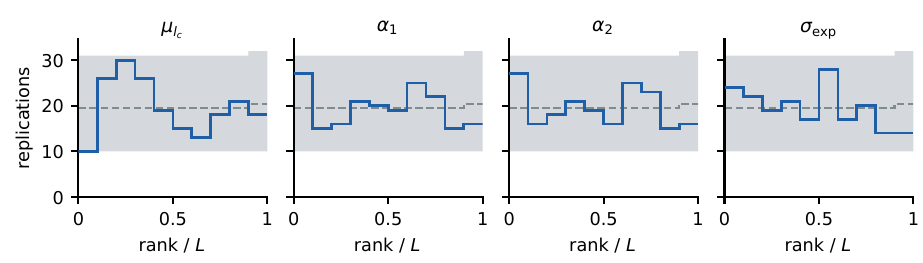}
 \caption{Rank histograms at the validation design for the population mean $\mu_{l_c}$ of the capillary length, the contact angles and the noise scale, from the 196 replications that finish and pass the convergence gate.
 Each rank is divided by $L=200$ and binned into ten bins, the last of which holds one attainable rank more than the others.
 Solid line: observed count.
 Dashed line: expected count under uniformity.
 Shaded: the band in which a uniform histogram stays with probability at least 0.99 in each bin, from the quantiles of the binomial distribution.
 \Cref{fig:sm_sbcranks} shows every parameter.}
 \label{fig:sbcranks}
\end{figure}

We ran the study at the validation design, where the map agrees with direct sampling in the wetting properties (\cref{sec:synthetic}), and at the real-data design, the design of the measured radii.
At the validation design of \cref{sec:hierarchy}, we ran 200 replications with the map of \cref{sec:training}.
All but four finish within the cap, and every replication that finishes passes the convergence gate.
The four capped replications have the four smallest true noise scales of the study, and they change no verdict.
The rank histograms of all 13 parameters stay inside their bands in every bin (\cref{fig:sm_sbcranks}; \cref{fig:sbcranks} shows $\mu_{l_c}$, the contact angles and the noise scale), and neither rule finds a departure for any parameter.

At the real-data design, we ran two independent studies of 200 replications each, of which 195 and 188 pass the convergence gate.
In both studies, the contact angles, the capillary lengths and the noise scale pass both rules.
Of these parameters, $\mu_{l_c}$ comes closest to a departure: it has the smallest $p$-value of the validation study, 0.0087 in the ECDF test, and $p=0.004$ in the Kolmogorov--Smirnov test of one real-data study.
Under the chi-squared and Kolmogorov--Smirnov rule, one study departs in $r_k$ and the other in $\Delta V_k$.
The ECDF test finds the departure in $\Delta V_k$.
The departure in $r_k$ rests on the chi-squared test of one study, and the Kolmogorov--Smirnov and ECDF tests of that study and all three tests of the other study find none.

We attribute the departure in $\Delta V_k$ to the redraw rule.
The rule removes ground truths outside the domain, where $\Delta V_k$ is larger on average and the real-data prior puts the most mass of the three designs, and the surrogate posterior places mass there (\cref{sec:sampling}).
The true $\Delta V_k$ then sit below the posterior mean on average, and the posteriors are wider than calibration requires.
At the validation design, with almost no prior mass outside the domain, $\Delta V_k$ shows no departure.

After the study, we split the validation replications into thirds by their true noise scale.
In the middle third, with true $\sigma_\text{exp}$ from \qty{0.22}{\mm} to \qty{0.43}{\mm}, the map places the noise scale above the truth by about 0.4 posterior standard deviations on average.
Within that third, the Kolmogorov--Smirnov $p=0.011$ lies above the study's Bonferroni threshold.
The lower and upper thirds show no such shift.
The synthetic ground truths of \cref{sec:synthetic}, at \qty{0.4}{\mm}, lie in the middle third.
In all three studies, the central \qty{90}{\percent} credible intervals of the noise scale cover the true value close to their nominal level.

With one map for all 200 posteriors, the validation study needs about five orders of magnitude fewer forward-model evaluations than direct sampling, with chains priced as in \cref{sec:cost}.
The amortized map thus makes simulation-based calibration of this hierarchical model affordable, and at the validation design the study detects no departure from calibration.
At the real-data design, every departure lies in the knife-edge and volume parameters.

\section{Real-data demonstration}\label{sec:realdata}

\subsection{Setup}\label{sec:realsetup}
We apply the amortized transport-map likelihood with the Young--Laplace forward model to the radii measured in the laboratory experiments of \cref{sec:experiment}, with direct sampling as the reference.
The data are six cells: three substrates, each at two volume settings (\cref{fig:experimentaldata}(a)).
Direct sampling and the map condition on these six radii.
The radii inform the fifteen parameters of the real-data design of \cref{sec:hierarchy}.
\Cref{sm:cells} lists the cells, and \cref{sm:tables} gives the values of this section for every parameter and map.

\subsection{The posterior conditioned on the real photographs}\label{sec:realposterior}
\begin{figure}[tbp]
 \centering
 {\small (a) posterior marginals}\par
 \includegraphics[width=\textwidth]{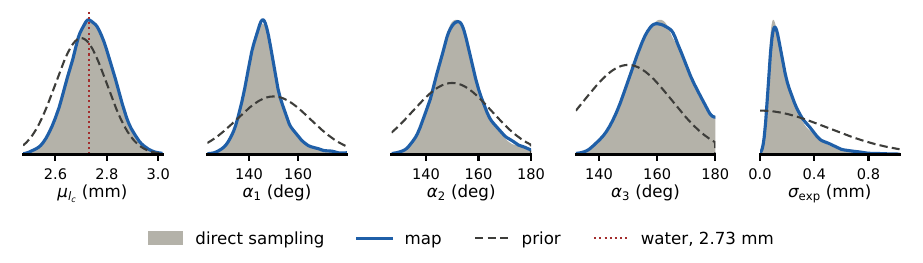}\par\medskip
 \begin{minipage}[t]{0.49\textwidth}\centering
 {\small (b) posterior predictive check}\par
 \includegraphics{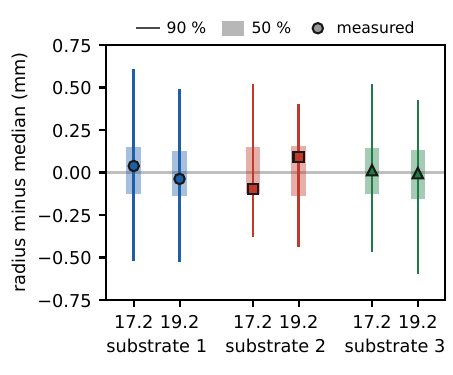}
 \end{minipage}\hfill
 \begin{minipage}[t]{0.49\textwidth}\centering
 {\small (c) agreement with direct sampling}\par
 \includegraphics{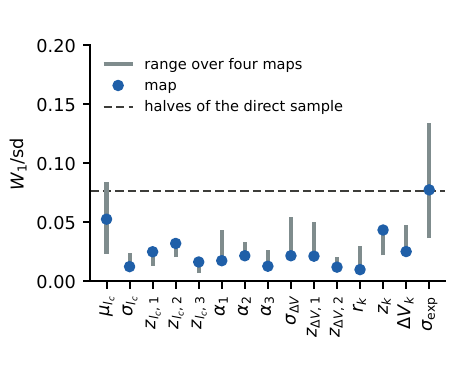}
 \end{minipage}
 \caption{The map on the real photographs.
 (a)~Posterior marginals of the population mean $\mu_{l_c}$ of the capillary length, the three contact angles and the noise scale.
 \Cref{sm:tables} shows every parameter.
 Shaded: direct sampling.
 Solid line: the map.
 Dashed line: the prior.
 Dotted line: the capillary length of water at \qty{20}{\celsius}.
 (b)~Posterior predictive check, one column per cell, labelled by substrate and nominal volume $V_{0j}$ in \unit{\ml}.
 Each cell is drawn about the median of its predictive radii.
 Line: central \qty{90}{\percent} interval.
 Bar: central \qty{50}{\percent} interval.
 Marker: the measured radius, in the shape and colour of \cref{fig:experimentaldata}(a).
 The intervals use every draw at which the reduced-order model returns a radius, 491 to 500 per cell.
 (c)~$W_1/\mathrm{sd}$ against direct sampling for every parameter, as in \cref{fig:agreement}(b).
 Marker: the map.
 Bar: the range over the map and its three replicate maps.
 Dashed line: $W_1/\mathrm{sd}$ between the two halves of the direct sample, maximized over the parameters.}
 \label{fig:realdata}
\end{figure}
We compare the posteriors of the map and of its replicate maps with that of direct sampling by $W_1/\mathrm{sd}$, as at the synthetic designs.
\Cref{fig:realdata}(a) overlays the marginals of the map on those of direct sampling for the population mean $\mu_{l_c}$ of the capillary length, the three contact angles and the noise scale.
They agree closely for every parameter, the noise scale included.
Three replicate maps, each trained on an independent training draw as in \cref{sec:synthetic}, agree closely too.
Over the four maps, $W_1/\mathrm{sd}$ stays below one seventh for the noise scale and below one tenth for every other parameter (\cref{fig:realdata}(c)).
It also stays below one tenth along every principal direction of the direct posterior in the contact angles and capillary lengths.
The correlation coefficients of all 105 pairs of parameters differ from those of direct sampling by at most 0.05 (\cref{fig:sm_triangle_real}).
For the map, $W_1/\mathrm{sd}$ is at most about the split-half level for every parameter, and the noise scale reaches it (\cref{fig:realdata}(c)).
The noise scale agrees more closely here than at the synthetic designs partly because six radii leave its posterior wide, with a standard deviation under direct sampling of \qty{0.15}{\mm} against \qty{0.06}{\mm} and \qty{0.02}{\mm} there.
The map shifts its posterior mean less than at the validation design, by about \qty{0.01}{\mm}.

The two volume settings of this design are the pair of \cref{fig:rankone} whose curves nearly coincide, so each substrate's two radii inform about one direction in contact angle and capillary length.
Under the map, the six radii narrow the contact angles to about half to two thirds of their prior standard deviation.
The capillary lengths stay above four fifths of theirs.
The population mean of the capillary length is $\mu_{l_c}=\qty[uncertainty-mode=separate]{2.740\pm0.086}{\mm}$ (posterior mean and standard deviation), consistent with the capillary length of about \qty{2.75}{\mm} that the measured surface tension of the dispersion gives (\cref{sec:priors}) and with the standard value for water, \qty{2.73}{\mm} at \qty{20}{\celsius}~\cite{IAPWS2014-st,IAPWS2011-lw}.
Because the prior is centred at \qty{2.7}{\mm}, near both values, this agreement is a consistency check.
On the real photographs, the map thus agrees closely with direct sampling, and the photographs inform the contact angles more than the capillary lengths.

\subsection{Posterior predictive check}\label{sec:ppc}
A posterior predictive check tests whether the posterior reproduces the measured radii.
For 500 draws of the map's posterior, the reduced-order model solves each cell, and \cref{eqn:observation} adds noise at the noise scale of the draw.
This gives each cell a posterior predictive distribution of its radius.
All six measured radii lie inside the central \qty{50}{\percent} and \qty{90}{\percent} intervals of their own cell (\cref{fig:realdata}(b)).
With six cells, this in-sample check has little power.

Sampling the map's posterior needs no forward solve, but this check runs the reduced-order model again after training.
Of its \num{3000} solves, 27 return no radius.
In each, the draw places a cell of the \qty{19.2}{\ml} setting beyond the domain bound of \cref{sec:rom}, where the surrogate posterior places mass (\cref{sec:sampling}).
\Cref{sec:sbcresults} attributes the departure in $\Delta V_k$ to the same mass.
From training through the posterior to the predictive check, the approach runs end to end on real photographs, and its posterior agrees with direct sampling and is consistent with the measured radii.
\FloatBarrier

\section{Conclusion}\label{sec:conclusion}
We introduced a framework for simulation-based inference in a hierarchical inverse problem, with one amortized triangular transport-map likelihood per design that serves every cell of the hierarchical model.
We applied it to infer the contact angle and the capillary length from photographs of a hole in a thin liquid film.
The map is trained offline on forward solves.
We tested the map against direct sampling of the forward model and by simulation-based calibration at synthetic designs, and applied it to experimentally measured radii.

In the present problem direct sampling of the forward model stays feasible and served as the reference.
The map's posteriors agree with direct sampling at the validation design, apart from an upward shift of the noise scale, and there the map also passes simulation-based calibration for every parameter.
We conclude that at the validation design the amortized map infers the wetting properties and their uncertainty accurately.
On the larger hierarchy of the scaling design the posteriors agree less closely: the map and its replicate maps shift the posterior mainly along one combination of contact angles and capillary lengths.
We hypothesize that the errors of the learned likelihood add up over the larger number of cells while the posterior narrows, similar to the compounding approximation errors reported for compositional amortized inference~\cite{Arruda2025-cq}.
The Gaussian special case shifts less along this combination at this design, so the learned width and shape may contribute.
A higher order of the map or more training rows may reduce this shift.
The learned width and shape of the noise account for most of the shift of the noise scale on Gaussian data, and under the skewed law we tested they bring the noise scale closer to direct sampling than the Gaussian special case does.
On real photographs the method runs end to end, from training through the posterior to the posterior predictive check, and the map agrees with direct sampling for every parameter, the noise scale included.
The map can place posterior mass at inputs where the reduced-order model returns no radius, and we attribute to this the departure of the calibration of $\Delta V_k$ in one of the two calibration studies at the real-data design.
A surrogate that also learns where the reduced-order model returns a radius would confine the posterior to the domain of the forward model.

Even for a single posterior at the validation design, the map needs orders of magnitude fewer forward-model evaluations than direct sampling with a gradient-free ensemble sampler, and once it is trained offline, a further posterior at its design needs no forward solve.
Studies that need many posteriors, such as calibration studies or repeated measurements, therefore take far fewer forward-model evaluations than direct sampling.
For an expensive forward model, forward solves dominate the running time, so the training draw accounts for most of the cost of the map, and the online cost of each posterior falls to that of sampling the surrogate.

The training draw itself may also shrink.
At the validation design, a map trained on a fraction of the forward solves agrees with direct sampling within the range of the full-budget maps, from one map per budget.
Training rows drawn near the posterior, as in sequential neural likelihood~\cite{Papamakarios2019-snl}, could lower the cost further for a single dataset.

Without direct sampling as a reference, simulation-based calibration can still check the map, and beyond the training draw it needs forward solves only to simulate its datasets.
The map thus makes this check affordable where direct sampling would make it costly.
Calibration and agreement with direct sampling test different properties: calibration averages over datasets drawn from the prior, and agreement compares the posteriors for one dataset.
At the validation design, the noise scale passes calibration while its posterior sits above that of direct sampling.
A comparison with direct sampling, on a problem small enough to afford it, therefore remains part of validating such a map.

The Bayesian analysis also shows what photographs of a hole determine.
A photograph informs a single combination of contact angle and capillary length, and at the volume settings of the experiments the photographs of one substrate add little beyond it.
The contact angles are then fixed only through the shared liquid and the prior on the capillary length.
An experiment designed for inference would choose volume settings at which the radius responds to contact angle and capillary length in more different proportions.
With such settings the same measurement would also give the surface tension of a liquid of known density.

The construction uses the forward model only through its solves, so it could also serve other hierarchical inverse problems with expensive forward models, provided that the likelihood factorizes over cells, every cell shares an observation density free of the hierarchy's hyperparameters, and a single parameter scales the simulator's noise exactly.
We will apply the construction to forward models of more complicated wetting processes, such as impacting drops~\cite{Josserand2016-di}.
A map trained offline for suitable settings could act as a soft sensor in a running process, such as coating or inkjet printing.
It would turn each photograph taken during the process into the contact angle and the surface tension with their uncertainty, without a forward solve and without disturbing the liquid.

\section*{Code availability}
A basic implementation of the method and of its application to the Young--Laplace problem will be made publicly available upon publication.

\section*{Acknowledgments}
The author thanks Youssef Marzouk and his group for discussions on Bayesian inference and transport maps.
The author used Anthropic's Claude (Opus 5.5, through Claude Code) to edit prose, to search the literature and to write the scripts that recompute the reported values and figures.
The author assumes responsibility for all content.

\bibliography{literature}

\clearpage
\begin{center}{\large\bfseries Supplementary Materials}\end{center}
\makeatletter
\@removefromreset{figure}{section}\@removefromreset{table}{section}
\renewcommand\thefigure{SM\@arabic\c@figure}
\renewcommand\thetable{SM\@arabic\c@table}
\renewcommand\thesection{SM\arabic{section}}
\makeatother
\renewcommand\thepage{SM\arabic{page}}
\setcounter{section}{0}\setcounter{figure}{0}\setcounter{table}{0}\setcounter{equation}{0}\setcounter{page}{1}
\section{The reduced-order model}\label{sm:rom}
The basis comes from \num{2220} full boundary-value solutions on a grid of five contact angles from \qtyrange{90}{180}{\degree}, five capillary lengths from \qtyrange{2.3}{3.1}{\mm} and the knife-edge radii \qtylist{30;40;50}{\mm}, with the knife-edge height and the volume varied across the solutions.
The knife-edge height is 0.10, 0.13 or 0.16 times the knife-edge radius.
The prior of \cref{tab:priors} stays within the band these ratios span, and the training distribution of \cref{sec:training} leaves it in the tails of $z_k$.

The accuracy check draws 50 parameter vectors from the hierarchical prior at each of four volumes, fixes the volume exactly, and solves twice: with the reduced-order model, and with the full boundary-value solver, whose contact-line radius $r(0)$ is root-found to meet the same volume.
A second check repeats this with the other inputs drawn from the training distribution, whose knife-edge coordinates are broader than the prior's.
Of its \num{204} draws, the reduced-order solve returns no radius at 4, which are redrawn.
Of the other 200, 48 have $z_k/r_k$ outside the band of the basis.
A third check draws 50 rows from the training distribution at each of five volumes from \qty{19.6}{\ml} to \qty{21.5}{\ml}.
Of its \num{341} draws, the reduced-order solve returns no radius at 77 and the full solve fails at 14, and these are redrawn.
\Cref{tab:sm_rom} gives the relative difference in $R$ for all three.

\begin{table}[tbhp]
 \centering
 \caption{Relative difference in $R$ between the reduced-order model and the full boundary-value solve off the training grid, per volume, with the other inputs drawn from the prior and from the training distribution.
 The volumes without prior entries are those of the third check.
 The last two rows split the training-distribution draws of the four volumes of the second check by whether $z_k/r_k$ lies in the band of the basis.}
 \label{tab:sm_rom}
 \begin{tabular}{@{}lrrrrrr@{}}
\toprule
& \multicolumn{3}{c}{prior} & \multicolumn{3}{c}{training distribution} \\
$V$ [mL] & draws & median [\%] & max [\%] & draws & median [\%] & max [\%] \\
\midrule
10.00 & 50 & 0.007 & 0.013 & 50 & 0.008 & 0.015 \\
17.18 & 50 & 0.054 & 0.157 & 50 & 0.053 & 0.248 \\
18.00 & 50 & 0.087 & 0.192 & 50 & 0.082 & 0.672 \\
19.21 & 50 & 0.115 & 0.544 & 50 & 0.141 & 1.049 \\
19.60 & -- & -- & -- & 50 & 0.175 & 2.759 \\
20.00 & -- & -- & -- & 50 & 0.171 & 2.754 \\
20.50 & -- & -- & -- & 50 & 0.208 & 1.860 \\
21.00 & -- & -- & -- & 50 & 0.266 & 2.597 \\
21.50 & -- & -- & -- & 50 & 0.231 & 1.654 \\
\midrule
10.00 to 19.21 & 200 & 0.066 & 0.544 & 200 & 0.062 & 1.049 \\
19.60 to 21.50 & & & & 250 & 0.202 & 2.759 \\
$z_k/r_k$ in the basis & & & & 152 & 0.057 & 1.049 \\
$z_k/r_k$ outside & & & & 48 & 0.084 & 0.495 \\
\bottomrule
\end{tabular}

\end{table}

\Cref{tab:sm_romconv} places the volume bound of \cref{sec:rom} against the failures of the production solve.
For this table, the production solve runs without the bound at \num{300} draws of the other inputs from the hierarchical prior and 15 volumes from \qtyrange{10}{24}{\ml}, and each entry is the share of draws for which it returns a radius.

\begin{table}[tbhp]
 \centering
 \caption{Share of prior draws for which the reduced-order solve returns a radius, per volume, without the volume bound.}
 \label{tab:sm_romconv}
 \begin{tabular}{@{}rrr@{}}
\toprule
$V$ [mL] & radius returned & share \\
\midrule
10.00 & 300/300 & 1.000 \\
12.00 & 300/300 & 1.000 \\
14.00 & 300/300 & 1.000 \\
16.00 & 300/300 & 1.000 \\
17.18 & 300/300 & 1.000 \\
18.00 & 300/300 & 1.000 \\
18.80 & 300/300 & 1.000 \\
19.21 & 298/300 & 0.993 \\
19.50 & 296/300 & 0.987 \\
20.00 & 292/300 & 0.973 \\
20.50 & 282/300 & 0.940 \\
21.00 & 264/300 & 0.880 \\
22.00 & 182/300 & 0.607 \\
23.00 & 78/300 & 0.260 \\
24.00 & 14/300 & 0.047 \\
\bottomrule
\end{tabular}

\end{table}

\section{Information in the photographs}\label{sm:information}
\Cref{tab:sm_information} gives the numbers behind \cref{sec:information}, from the reduced-order grids of \cref{fig:rankone} at the reference point, with $\alpha$ in units of \qty{15}{\degree} and $l_c$ in units of \qty{0.1}{\mm}.
For two photographs of one substrate, the ratio of the standard deviations along and across the curve is the square root of the ratio of the largest to the smallest eigenvalue of their information.
For three substrates with two photographs each at the real-data volumes, a direction of the information counts as weak when its eigenvalue lies below $10^{-3}$ of the largest.

\begin{table}[tbhp]
 \centering
 \caption{Information of the photographs in contact angle and capillary length: standard deviations along and across the curve of \cref{fig:rankone} for two photographs, and weak directions of the information of three substrates, with the smallest eigenvalue and the next one, each relative to the largest.}
 \label{tab:sm_information}
 \begin{tabular}{@{}lll@{}}
\toprule
case & quantity & value \\
\midrule
\qty{17.2}{\ml} and \qty{19.2}{\ml} (real data) & sd along / sd across & 273 \\
\qty{10}{\ml} and \qty{18}{\ml} (validation) & sd along / sd across & 46 \\
\midrule
own capillary length & weak directions & 3 (each $1.3\times10^{-5}$; next $1.00$) \\
shared capillary length & weak directions & 1 ($1.2\times10^{-5}$; next $0.31$) \\
\bottomrule
\end{tabular}

\end{table}

\section{Software and settings}\label{sm:software}
All computations ran in Python 3.12.7 with NumPy 2.4.6, SciPy 1.15.3, MParT 2.2.2, emcee 3.1.6 and littlemcmc 0.2.2.

\paragraph{Likelihood component}
The mean $a$ is a probabilists' Hermite expansion of total order three in the five inputs $\bm t$ and has 56 coefficients.
The correction $d$ has total order two in the six inputs $\bm x$ and 28 coefficients.
The monotone component $B$ has total order three in the seven inputs $(\bm x,r)$ and 120 coefficients.
Each map of \cref{tab:sm_maps} therefore has 204 coefficients.

\paragraph{Training}
Each row of $q_{\bm\theta}$ is one cell, at a substrate and a volume setting chosen at random, of a draw from the hierarchical prior with two changes.
The knife-edge coordinates $r_k$ and $z_k$ have standard deviation \qty{1}{\mm}, against \qty{0.333}{\mm} in the prior, and the volume error has an additional population mean drawn from $\mathcal N(0,\ \qty{0.5}{\ml})$.
The mixture $q_\sigma$ draws $\sigma_\text{exp}$ with equal probability from the half-normal prior of \cref{tab:priors} and from a log-uniform distribution on \qtyrange{0.002}{2}{\mm}.
The ridge penalty is $10^{-6}$ times the squared norm of the coefficients.
L-BFGS-B runs for at most \num{20000} iterations.
\Cref{tab:sm_maps} lists the trained maps with their forward solves.
The analytic log-likelihood gradient agrees with central finite differences to a relative error of $8.8\times10^{-7}$ on 40 training rows, and the gradient of \cref{eqn:logpost} to $9.8\times10^{-6}$ at three prior draws in eight coordinates.

\begin{table}[tbhp]
 \centering
 \caption{The trained maps by family.
 Rows with a radius: the rows of the training draw below the \qty{20}{\ml} bound of \cref{sec:rom} at which the reduced-order model returns a radius.
 The draws of the three maps of \cref{sec:training} take \num{1933}, \num{1990} and \num{1999} forward solves at the real-data, validation and scaling designs.
 The replicate maps train on independent training draws of fresh parameter rows (\cref{sec:synthetic,sec:realposterior}), each with as many rows with a radius as the map of its design.
 The maps with $N$ train on the first $N$ rows with a radius of a random permutation of the map draw of their design.}
 \label{tab:sm_maps}
 \begin{tabular}{@{}lrrr@{}}
\toprule
 & \multicolumn{3}{c}{rows with a radius} \\ \cmidrule(lr){2-4}
map & validation & scaling & real data \\
\midrule
map & \num{1989} & \num{1997} & \num{1826} \\
replicate 1 & \num{1989} & \num{1997} & \num{1826} \\
replicate 2 & \num{1989} & \num{1997} & \num{1826} \\
replicate 3 & \num{1989} & \num{1997} & \num{1826} \\
$N=\num{250}$ & \num{250} & -- & \num{250} \\
$N=\num{500}$ & \num{500} & -- & \num{500} \\
$N=\num{1000}$ & \num{1000} & -- & \num{1000} \\
\bottomrule
\end{tabular}

\end{table}

\paragraph{Sampling}
The No-U-Turn sampler runs four chains of \num{4000} warm-up and \num{10000} draws at a target acceptance rate of 0.95.
In simulation-based calibration each replication runs four chains of \num{3000} warm-up and \num{2000} draws.
A replication that has not finished after \qty{4}{\hour} is stopped and run once more with a cap of \qty{16}{\hour}.

Direct sampling uses the ensemble sampler of \cref{sec:goldstandard} with 40 walkers and the default stretch move.
Its chains run for \num{250000} steps at the validation and scaling designs and \num{120000} at the real-data design, and the comparisons use their last \num{95000}, \num{100000} and \num{55000} steps, thinned by 50.
The direct chain of the skewed-noise test runs for \num{40000} steps after \num{1000} burn-in steps.

\FloatBarrier

\section{Convergence}\label{sm:convergence}
\Cref{tab:sm_conv} gives the largest $\hat R$ and the smallest bulk effective sample size~\cite{Vehtari2021-rhat}.
For the ensemble sampler, each walker counts as a chain.
$\hat R$ assumes independent chains, and the walkers interact.
The direct chain at the real-data design comes closest to the limit of \cref{sec:sampling}, with $\hat R=1.048$, and its marginals agree with those of the map's independent No-U-Turn chains to about the split-half level of \cref{sec:goldstandard} (\cref{sec:realposterior}).
In simulation-based calibration, every replication that passes the convergence gate of \cref{sec:sbc} also stays within this limit, with a largest $\hat R$ of 1.03.

\begin{table}[tbhp]
 \centering
 \caption{Convergence of every chain behind \cref{sec:agreement,sec:realdata}: largest rank-normalized split $\hat R$ and smallest bulk effective sample size over the parameters and, for the maps, over every map sampled at that design (the maps of \cref{tab:sm_maps}, the two Gaussian forms and, for skewed noise, the three components of \cref{tab:sm_skew}), each with four chains of the No-U-Turn sampler.
 Direct chains: the steps of \cref{sm:software} that the comparisons use.}
 \label{tab:sm_conv}
 \begin{tabular}{@{}lllrr@{}}
\toprule
design & chains & sampler & max $\hat R$ & min bulk ESS \\
\midrule
validation & 9 maps & map & 1.00 & \num{8858} \\
scaling & 6 maps & map & 1.00 & \num{10259} \\
real data & 7 maps & map & 1.00 & \num{2535} \\
validation, skewed noise & 3 maps & map & 1.00 & \num{15102} \\
\midrule
validation & 40 walkers & direct & 1.009 & \num{3605} \\
scaling & 40 walkers & direct & 1.039 & \num{739} \\
real data & 40 walkers & direct & 1.048 & \num{684} \\
validation, skewed noise & 40 walkers & direct & 1.040 & \num{804} \\
\bottomrule
\end{tabular}

\end{table}

Two checks test the direct sampler against known answers.
The first uses a sentinel coordinate, which has a standard normal prior and enters no likelihood and therefore has posterior standard deviation one.
The second uses a flat direction.
By \cref{eqn:noncentred}, adding a constant $c$ to every $z_{\Delta V,j}$ and $-\sigma_{\Delta V}c$ to $\Delta V_k$ leaves every $\Delta V_j$ unchanged, so the likelihood is constant along this direction.
With $J$ volume settings, $\bar z$ the mean of the $z_{\Delta V,j}$ and $\bar{\Delta V}$ that of the $\Delta V_j$, the prior terms that depend on $\bar z$ at fixed $\Delta V_j$ are $-J\bar z^{\,2}/2-(\bar{\Delta V}-\sigma_{\Delta V}\bar z)^2/(2\cdot(\qty{0.5}{\ml})^2)$, with \qty{0.5}{\ml} the prior standard deviation of $\Delta V_k$.
Given the $\Delta V_j$ and $\sigma_{\Delta V}$, the posterior standard deviation of $\sum_j z_{\Delta V,j}/\sqrt J=\sqrt J\,\bar z$ is therefore
\begin{equation}
 P=\sqrt{\frac{J}{J+4\sigma_{\Delta V}^2/\unit{\ml\squared}}}\;,
\end{equation}
which tends to the prior value one as $\sigma_{\Delta V}\to0$.
The marginal posterior standard deviation that a chain measures also contains the spread of this conditional mean over the posterior of the $\Delta V_j$ and $\sigma_{\Delta V}$, which can move it away from $P$.
We evaluate $P$ at each chain's posterior mean of $\sigma_{\Delta V}$, and the criterion, fixed before the measurement, is agreement within \qty{3}{\percent}.
\Cref{tab:sm_checks} gives both checks.

\begin{table}[tbhp]
 \centering
 \caption{The two checks of the direct sampler on the steps the comparisons use.
 Flat direction: measured and predicted posterior standard deviation of $\sum_j z_{\Delta V,j}/\sqrt J$ and their ratio.
 Sentinel: posterior standard deviation of a coordinate whose posterior is its standard normal prior.}
 \label{tab:sm_checks}
 \begin{tabular}{@{}lrrrr@{}}
\toprule
 & \multicolumn{3}{c}{flat direction} & \\ \cmidrule(lr){2-4}
design & measured & predicted & ratio & sentinel sd \\
\midrule
validation & 0.982 & 0.963 & 1.019 & 0.994 \\
scaling & 0.998 & 0.999 & 0.999 & 0.974 \\
real data & 0.966 & 0.945 & 1.022 & 0.989 \\
\bottomrule
\end{tabular}

\end{table}

\FloatBarrier

\section{Agreement, cost and the real-data posterior}\label{sm:tables}
The agreement study covers one synthetic dataset per design, drawn at the ground truth of \cref{tab:sm_truths}.

\begin{table}[tbhp]
 \centering
 \caption{Ground truths of the two synthetic designs.}
 \label{tab:sm_truths}
 \begin{tabular}{@{}llll@{}}
\toprule
parameter & unit & validation & scaling \\
\midrule
$\mu_{l_c}$ & mm & 2.72 & 2.72 \\
$\sigma_{l_c}$ & mm & 0.03 & 0.03 \\
$z_{l_c,1}$ & -- & $-0.3$ & $-0.3$ \\
$z_{l_c,2}$ & -- & 0.3 & 0 \\
$z_{l_c,3}$ & -- & -- & 0.3 \\
$\alpha_{1}$ & \unit{\degree} & 145 & 145 \\
$\alpha_{2}$ & \unit{\degree} & 152 & 148.5 \\
$\alpha_{3}$ & \unit{\degree} & -- & 152 \\
$\sigma_{\Delta V}$ & mL & 0.15 & 0.15 \\
$z_{\Delta V,1}$ & -- & $-0.5$ & $-0.5$ \\
$z_{\Delta V,2}$ & -- & 0.5 & $-0.3$ \\
$z_{\Delta V,3}$ & -- & -- & $-0.1$ \\
$z_{\Delta V,4}$ & -- & -- & 0.1 \\
$z_{\Delta V,5}$ & -- & -- & 0.3 \\
$z_{\Delta V,6}$ & -- & -- & 0.5 \\
$r_k$ & mm & 40 & 40 \\
$z_k$ & mm & 5 & 5 \\
$\Delta V_k$ & mL & 0.2 & 0.2 \\
$\sigma_\text{exp}$ & mm & 0.4 & 0.4 \\
\bottomrule
\end{tabular}

\end{table}

\begin{figure}[tbhp]
 \centering
 \includegraphics[width=\textwidth]{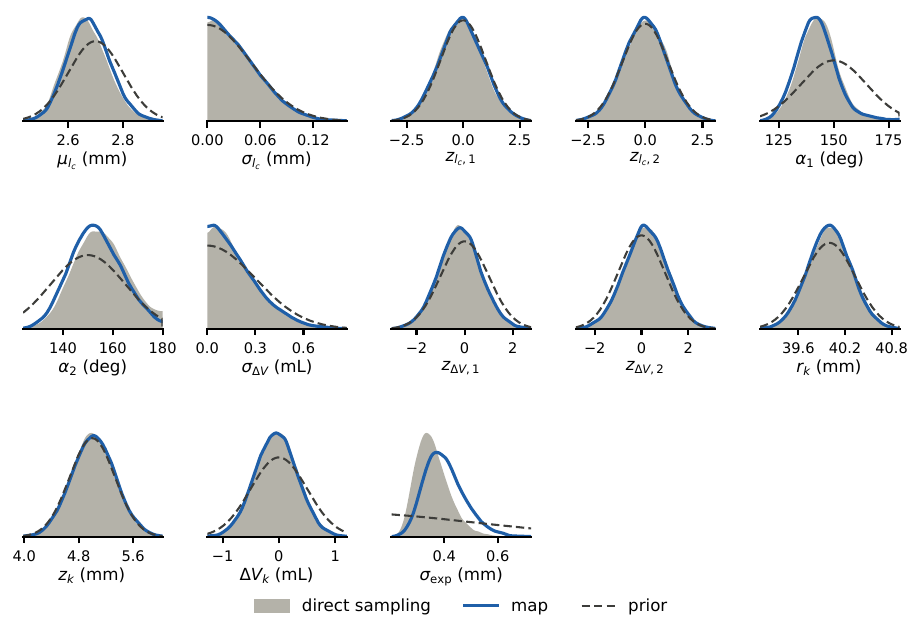}
 \caption{Posterior marginals at the validation design, one panel per parameter.
 Shaded: direct sampling.
 Solid line: the map.
 Dashed line: the prior.}
 \label{fig:sm_overlays_valid}
\end{figure}

\begin{figure}[tbhp]
 \centering
 \includegraphics[width=\textwidth]{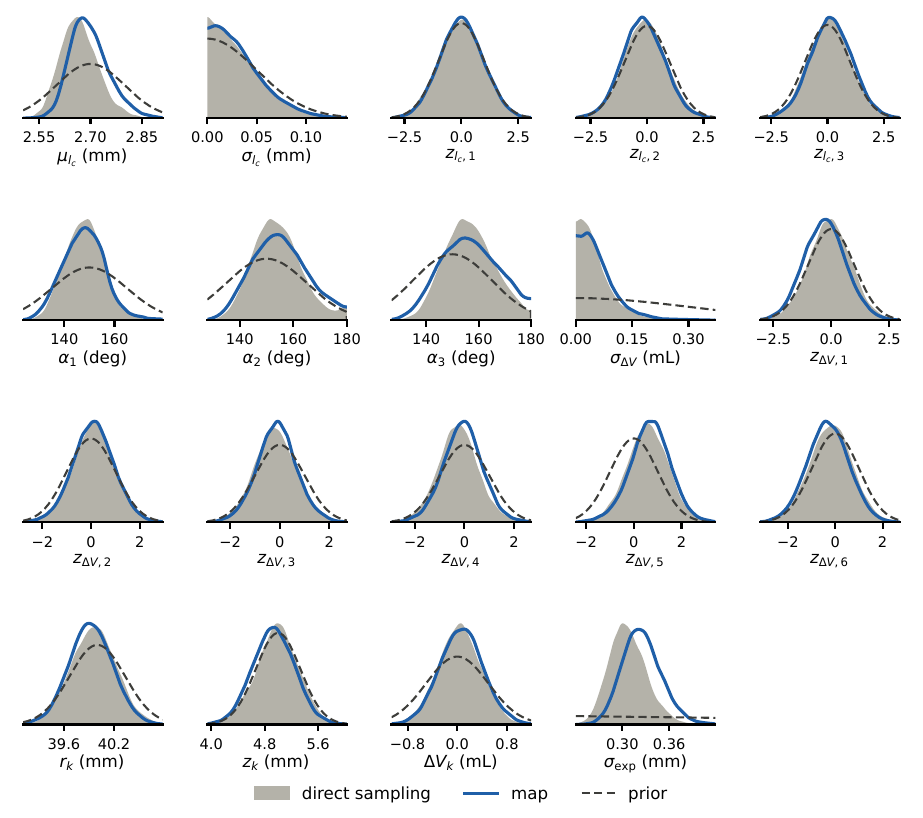}
 \caption{Posterior marginals at the scaling design, one panel per parameter.
 Shaded: direct sampling.
 Solid line: the map.
 Dashed line: the prior.}
 \label{fig:sm_overlays_scale}
\end{figure}

\Cref{fig:sm_overlays_valid,fig:sm_overlays_scale} overlay the marginals of the map on those of direct sampling for every parameter at the validation and scaling designs.
\Cref{tab:sm_w1sd_valid,tab:sm_w1sd_scale} give $W_1/\mathrm{sd}$ for the four learned maps, the Gaussian special case of \cref{sec:noisemodel} and the same form with its mean refitted by the training objective, with the posterior mean of the noise scale under each.
Refitting the mean leaves the noise scale of the Gaussian special case unchanged ($W_1/\mathrm{sd}$ of 0.17 and 0.17 at the validation design, 0.04 and 0.05 at the scaling design).
The last column is the $W_1/\mathrm{sd}$ between the two halves of the direct sample, and its largest value, the split-half level, is the dashed line of \cref{fig:agreement}(b).
The map and its three replicate maps spread most in the noise scale at the validation design, where the map lies at the top of its range, and in the contact angles, $\mu_{l_c}$ and the noise scale at the scaling design.
Against the split-half level, the map differs from direct sampling at the validation design by at most about 3.4 times this level for every parameter other than the noise scale.
At the scaling design it differs by at most about twice this level for every parameter other than the noise scale and $\mu_{l_c}$, which lies at about five times.
The noise scale differs by about 14 times this level at the validation design and 6.5 times at the scaling design.

\begin{table}[tbhp]
 \centering
 \caption{$W_1/\mathrm{sd}$ against direct sampling at the validation design.
 Learned component: the map and its three replicate maps (rep.~1 to 3).
 Gaussian: the special case $d=0$, $B$ the identity, with $a$ fitted by least squares, and the same form with $a$ refitted by the training objective (mean refitted).
 Last row: posterior mean $\bar\sigma_\text{exp}$ of the noise scale under each map, and under direct sampling in the last column.}
 \label{tab:sm_w1sd_valid}
 \begin{tabular}{@{}lrrrrrrr@{}}
\toprule
 & \multicolumn{4}{c}{learned} & \multicolumn{2}{c}{Gaussian} & direct \\ \cmidrule(lr){2-5}\cmidrule(lr){6-7}\cmidrule(lr){8-8}
parameter & map & rep.~1 & rep.~2 & rep.~3 & special case & mean refitted & split half \\
\midrule
$\mu_{l_c}$ & 0.13 & 0.02 & 0.16 & 0.14 & 0.05 & 0.03 & 0.014 \\
$\sigma_{l_c}$ & 0.02 & 0.01 & 0.02 & 0.01 & 0.05 & 0.01 & 0.008 \\
$z_{l_c,1}$ & 0.06 & 0.01 & 0.01 & 0.09 & 0.06 & 0.01 & 0.012 \\
$z_{l_c,2}$ & 0.04 & 0.02 & 0.07 & 0.06 & 0.04 & 0.02 & 0.009 \\
$\alpha_{1}$ & 0.14 & 0.08 & 0.01 & 0.17 & 0.08 & 0.06 & 0.038 \\
$\alpha_{2}$ & 0.16 & 0.01 & 0.04 & 0.19 & 0.12 & 0.15 & 0.048 \\
$\sigma_{\Delta V}$ & 0.01 & 0.01 & 0.05 & 0.03 & 0.01 & 0.07 & 0.013 \\
$z_{\Delta V,1}$ & 0.01 & 0.01 & 0.02 & 0.06 & 0.04 & 0.04 & 0.016 \\
$z_{\Delta V,2}$ & 0.03 & 0.05 & 0.08 & 0.03 & 0.07 & 0.02 & 0.048 \\
$r_k$ & 0.06 & 0.08 & 0.02 & 0.18 & 0.02 & 0.12 & 0.045 \\
$z_k$ & 0.04 & 0.01 & 0.15 & 0.02 & 0.03 & 0.04 & 0.019 \\
$\Delta V_k$ & 0.07 & 0.07 & 0.09 & 0.22 & 0.04 & 0.07 & 0.016 \\
$\sigma_\text{exp}$ & 0.68 & 0.28 & 0.34 & 0.45 & 0.17 & 0.17 & 0.021 \\
\midrule
$\bar\sigma_\text{exp}$ [mm] & 0.401 & 0.375 & 0.378 & 0.386 & 0.368 & 0.368 & 0.357 \\
\bottomrule
\end{tabular}

\end{table}

\begin{table}[tbhp]
 \centering
 \caption{As \cref{tab:sm_w1sd_valid}, at the scaling design.}
 \label{tab:sm_w1sd_scale}
 \begin{tabular}{@{}lrrrrrrr@{}}
\toprule
 & \multicolumn{4}{c}{learned} & \multicolumn{2}{c}{Gaussian} & direct \\ \cmidrule(lr){2-5}\cmidrule(lr){6-7}\cmidrule(lr){8-8}
parameter & map & rep.~1 & rep.~2 & rep.~3 & special case & mean refitted & split half \\
\midrule
$\mu_{l_c}$ & 0.54 & 0.04 & 0.10 & 0.22 & 0.02 & 0.07 & 0.067 \\
$\sigma_{l_c}$ & 0.03 & 0.05 & 0.05 & 0.02 & 0.04 & 0.02 & 0.047 \\
$z_{l_c,1}$ & 0.05 & 0.13 & 0.15 & 0.19 & 0.04 & 0.05 & 0.039 \\
$z_{l_c,2}$ & 0.10 & 0.05 & 0.03 & 0.08 & 0.01 & 0.06 & 0.021 \\
$z_{l_c,3}$ & 0.09 & 0.04 & 0.06 & 0.02 & 0.01 & 0.05 & 0.014 \\
$\alpha_{1}$ & 0.08 & 0.58 & 0.62 & 0.61 & 0.19 & 0.14 & 0.041 \\
$\alpha_{2}$ & 0.16 & 0.43 & 0.51 & 0.33 & 0.13 & 0.04 & 0.050 \\
$\alpha_{3}$ & 0.15 & 0.54 & 0.56 & 0.48 & 0.24 & 0.14 & 0.114 \\
$\sigma_{\Delta V}$ & 0.13 & 0.08 & 0.08 & 0.11 & 0.16 & 0.13 & 0.102 \\
$z_{\Delta V,1}$ & 0.24 & 0.23 & 0.29 & 0.25 & 0.16 & 0.23 & 0.040 \\
$z_{\Delta V,2}$ & 0.04 & 0.02 & 0.04 & 0.01 & 0.06 & 0.01 & 0.021 \\
$z_{\Delta V,3}$ & 0.10 & 0.07 & 0.09 & 0.06 & 0.10 & 0.16 & 0.056 \\
$z_{\Delta V,4}$ & 0.23 & 0.19 & 0.19 & 0.19 & 0.29 & 0.28 & 0.021 \\
$z_{\Delta V,5}$ & 0.13 & 0.09 & 0.06 & 0.10 & 0.13 & 0.08 & 0.014 \\
$z_{\Delta V,6}$ & 0.13 & 0.05 & 0.04 & 0.06 & 0.26 & 0.22 & 0.017 \\
$r_k$ & 0.14 & 0.20 & 0.12 & 0.13 & 0.04 & 0.02 & 0.025 \\
$z_k$ & 0.19 & 0.02 & 0.17 & 0.02 & 0.17 & 0.08 & 0.046 \\
$\Delta V_k$ & 0.10 & 0.13 & 0.04 & 0.01 & 0.04 & 0.01 & 0.011 \\
$\sigma_\text{exp}$ & 0.74 & 0.90 & 0.80 & 0.43 & 0.04 & 0.05 & 0.027 \\
\midrule
$\bar\sigma_\text{exp}$ [mm] & 0.325 & 0.329 & 0.327 & 0.318 & 0.309 & 0.309 & 0.307 \\
\bottomrule
\end{tabular}

\end{table}

\paragraph{Joint directions}
\Cref{tab:sm_joint} gives $W_1/\mathrm{sd}$ along the principal directions of the direct posterior in the contact angles and capillary lengths.
Their coordinates are $\alpha_i/\qty{15}{\degree}$ and $l_{c,i}/\qty{0.1}{\mm}$, with $l_{c,i}=\mu_{l_c}+\sigma_{l_c}z_{l_c,i}$, and the directions are the eigenvectors of the covariance of these coordinates under direct sampling.
At the scaling design, direction 4 has the components $(0.60, 0.22, 0.29, -0.02, 0.52, 0.49)$ in $(\alpha_1,\alpha_2,\alpha_3,l_{c,1},l_{c,2},l_{c,3})$.
All four maps place the posterior higher along it than direct sampling does, with the mean shifted by 0.48 to 0.82 direct posterior standard deviations and the standard deviation within ten percent of that of direct sampling.

\begin{table}[tbhp]
 \centering
 \caption{$W_1/\mathrm{sd}$ against direct sampling along the principal directions of the direct posterior in $(\alpha_i/\qty{15}{\degree}, l_{c,i}/\qty{0.1}{\mm})$, ordered by their standard deviation under direct sampling (sd, in these units).
 Replicate maps: replicates 1 to 3 of \cref{tab:sm_maps}.
 Gaussian: the special case and the same form with its mean refitted, as in \cref{tab:sm_w1sd_valid}, at the synthetic designs.
 Split half: the same between the two halves of the direct sample.}
 \label{tab:sm_joint}
 \begin{tabular}{@{}llrrrrrrrr@{}}
\toprule
design & direction & sd & map & \multicolumn{3}{c}{replicate maps} & \multicolumn{2}{c}{Gaussian} & split half \\
\midrule
validation & 1 & 1.22 & 0.16 & 0.01 & 0.15 & 0.18 & 0.05 & 0.04 & 0.015 \\
 & 2 & 0.68 & 0.07 & 0.04 & 0.10 & 0.11 & 0.05 & 0.13 & 0.052 \\
 & 3 & 0.53 & 0.03 & 0.06 & 0.08 & 0.05 & 0.18 & 0.03 & 0.012 \\
 & 4 & 0.15 & 0.15 & 0.17 & 0.05 & 0.21 & 0.23 & 0.07 & 0.009 \\
\midrule
scaling & 1 & 1.17 & 0.31 & 0.29 & 0.38 & 0.41 & 0.12 & 0.07 & 0.055 \\
 & 2 & 0.50 & 0.49 & 0.29 & 0.26 & 0.08 & 0.10 & 0.05 & 0.114 \\
 & 3 & 0.49 & 0.06 & 0.12 & 0.07 & 0.11 & 0.11 & 0.14 & 0.067 \\
 & 4 & 0.45 & 0.69 & 0.82 & 0.75 & 0.48 & 0.26 & 0.22 & 0.097 \\
 & 5 & 0.12 & 0.15 & 0.17 & 0.18 & 0.10 & 0.15 & 0.10 & 0.075 \\
 & 6 & 0.10 & 0.12 & 0.22 & 0.25 & 0.23 & 0.10 & 0.12 & 0.058 \\
\midrule
real data & 1 & 1.50 & 0.04 & 0.04 & 0.02 & 0.08 & -- & -- & 0.042 \\
 & 2 & 0.79 & 0.02 & 0.02 & 0.02 & 0.02 & -- & -- & 0.074 \\
 & 3 & 0.58 & 0.02 & 0.03 & 0.03 & 0.03 & -- & -- & 0.058 \\
 & 4 & 0.50 & 0.04 & 0.04 & 0.04 & 0.03 & -- & -- & 0.055 \\
 & 5 & 0.18 & 0.02 & 0.02 & 0.02 & 0.04 & -- & -- & 0.092 \\
 & 6 & 0.15 & 0.03 & 0.05 & 0.05 & 0.04 & -- & -- & 0.031 \\
\bottomrule
\end{tabular}

\end{table}

\paragraph{Four maps pooled}
Pooled with equal weights, the draws of the map and its three replicate maps give $W_1/\mathrm{sd}$ against direct sampling of 0.44, 0.72 and 0.06 in the noise scale at the validation, scaling and real-data designs.
Outside the noise scale they give at most 0.11, 0.45 and 0.05.

\begin{figure}[tbhp]
 \centering
 \includegraphics[width=\textwidth]{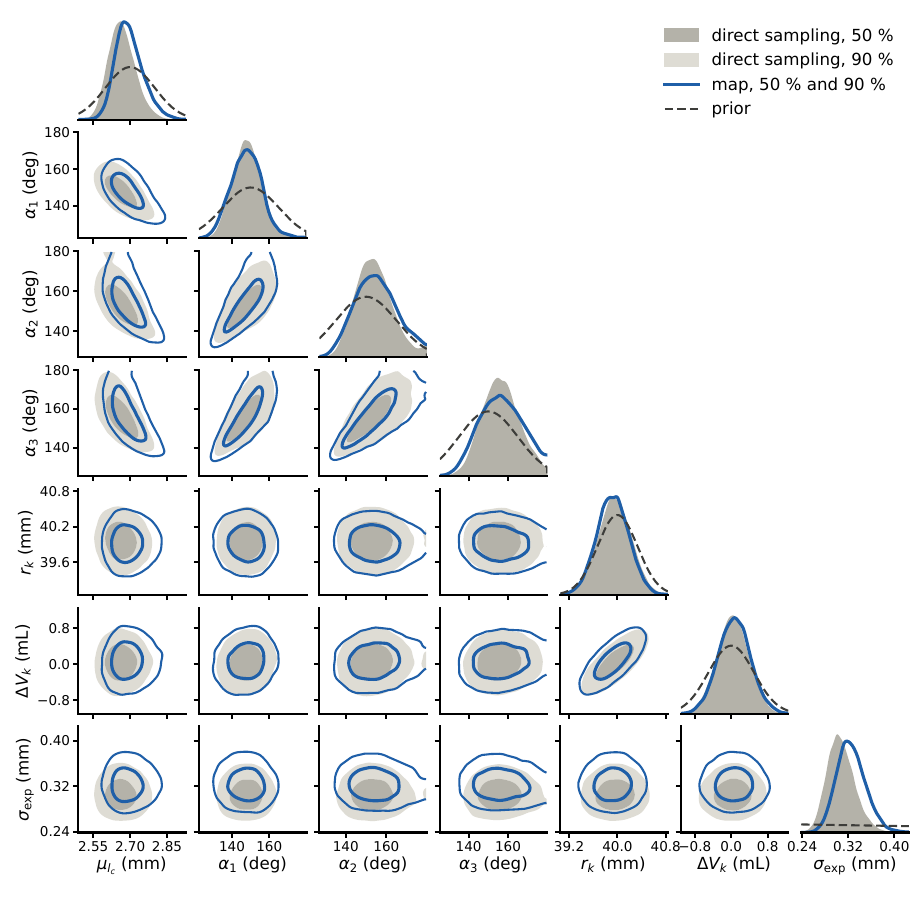}
 \caption{Pairwise joint posteriors at the scaling design, drawn as in \cref{fig:triangle}.}
 \label{fig:sm_triangle_scale}
\end{figure}

\begin{figure}[tbhp]
 \centering
 \includegraphics[width=\textwidth]{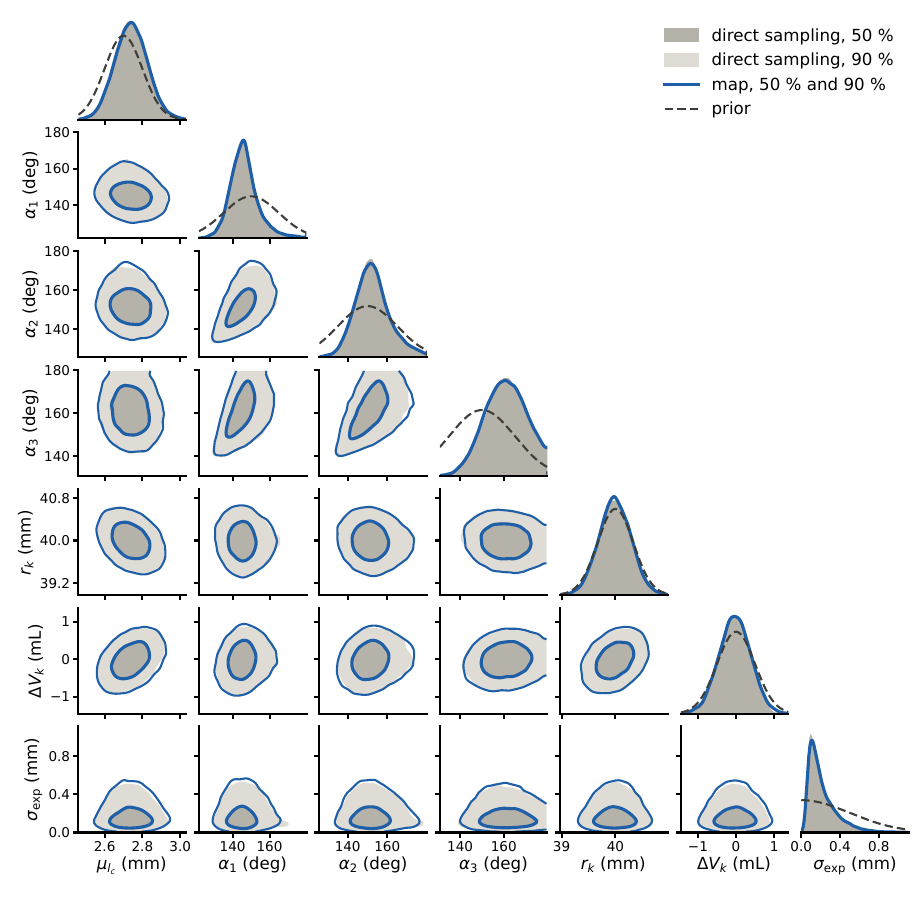}
 \caption{Pairwise joint posteriors at the real-data design, drawn as in \cref{fig:triangle}.}
 \label{fig:sm_triangle_real}
\end{figure}

\paragraph{Pairwise joint posteriors}
\Cref{fig:sm_triangle_scale,fig:sm_triangle_real} show the pairwise joint posteriors at the scaling and real-data designs, as \cref{fig:triangle} does at the validation design.
At the scaling design, the correlation coefficients of all 171 pairs of parameters differ from those of direct sampling by at most 0.09, with a median of 0.02.
The map couples the contact angles more strongly there, with coefficients of 0.66 to 0.67 against 0.60 to 0.61 under direct sampling.
A linear-Gaussian approximation from the gradients of the forward model at the direct posterior mean and the priors, for every parameter other than the noise scale, gives the sign of every correlation above 0.2 in magnitude, at all three designs, and correlates with the measured coefficients at 0.98, 0.93 and 0.95.

\paragraph{Training budget}
\Cref{tab:sm_budget} gives $W_1/\mathrm{sd}$ for the maps with $N=250$, 500 and \num{1000} of \cref{tab:sm_maps}, with all other settings unchanged.
The four maps of full budget at the validation design are in \cref{tab:sm_w1sd_valid}.

\begin{table}[tbhp]
 \centering
 \caption{$W_1/\mathrm{sd}$ against direct sampling for maps trained on 250, 500 and \num{1000} rows.}
 \label{tab:sm_budget}
 \begin{tabular}{@{}lrrrrrr@{}}
\toprule
 & \multicolumn{3}{c}{validation} & \multicolumn{3}{c}{real data} \\ \cmidrule(lr){2-4}\cmidrule(lr){5-7}
parameter & $250$ & $500$ & $\num{1000}$ & $250$ & $500$ & $\num{1000}$ \\
\midrule
$\mu_{l_c}$ & 0.13 & 0.05 & 0.05 & 0.12 & 0.06 & 0.05 \\
$\sigma_{l_c}$ & 0.01 & 0.01 & 0.01 & 0.01 & 0.01 & 0.02 \\
$z_{l_c,1}$ & 0.06 & 0.04 & 0.05 & 0.02 & 0.03 & 0.01 \\
$z_{l_c,2}$ & 0.02 & 0.01 & 0.06 & 0.01 & 0.01 & 0.05 \\
$z_{l_c,3}$ & -- & -- & -- & 0.01 & 0.01 & 0.02 \\
$\alpha_{1}$ & 0.16 & 0.10 & 0.05 & 0.03 & 0.03 & 0.02 \\
$\alpha_{2}$ & 0.09 & 0.01 & 0.09 & 0.01 & 0.02 & 0.06 \\
$\alpha_{3}$ & -- & -- & -- & 0.08 & 0.02 & 0.03 \\
$\sigma_{\Delta V}$ & 0.04 & 0.03 & 0.03 & 0.05 & 0.09 & 0.01 \\
$z_{\Delta V,1}$ & 0.01 & 0.06 & 0.01 & 0.03 & 0.04 & 0.02 \\
$z_{\Delta V,2}$ & 0.05 & 0.07 & 0.03 & 0.02 & 0.03 & 0.05 \\
$r_k$ & 0.06 & 0.06 & 0.02 & 0.03 & 0.03 & 0.02 \\
$z_k$ & 0.06 & 0.02 & 0.02 & 0.03 & 0.05 & 0.05 \\
$\Delta V_k$ & 0.07 & 0.02 & 0.02 & 0.09 & 0.03 & 0.02 \\
$\sigma_\text{exp}$ & 0.41 & 0.41 & 0.67 & 0.08 & 0.06 & 0.10 \\
\bottomrule
\end{tabular}

\end{table}

\paragraph{Mean of the likelihood component}
At 500 posterior draws of the map per design, spread over its four chains, the mean $\mu_y+s_ya(\bm t)$ differs from the reduced-order radius by at most 0.79, 0.60 and 0.59 times the draw's noise scale at the validation, scaling and real-data designs, in every cell but one.
That cell, at the validation design, is a lower root of the reduced-order solve, with a radius of \qty{2.0}{\mm}, and the mean differs from it by \qty{16.8}{\mm}.
The upper root, found by continuation in volume and confirmed by the full solver, is \qty{18.8}{\mm}, and the mean lies 0.2 noise scales from it.
Over the training rows the difference exceeds \qty{1}{\mm} at \qtylist{2.2;1.0;0.9}{\percent} of the rows, the largest of them at small radii.
The training rows of the twelve maps at full budget include, at radii of at most \qty{8}{\mm}, 0 to 4 such lower roots per map, at most \qty{0.22}{\percent} of the rows.
Direct sampling uses the same solve; a chain keeps no draw at a lower root, whose likelihood is lower by at least 125 nats.

\paragraph{Skewed noise}
The skewed law is $\varepsilon=(X-m)/s$ with $X=\sinh(\operatorname{arsinh}Z+e)$, $Z\sim\mathcal N(0,1)$ and $e=0.75$, the sinh-arcsinh law of Jones and Pewsey~\cite{Jones2009-sinh} with skewness parameter $e$ and tailweight parameter one, where $m$ and $s$ are the mean and standard deviation of $X$.
Its skewness is 1.004.
The dataset is drawn at the validation ground truth of \cref{tab:sm_truths} with this law in place of the Gaussian.
Over its 20 radii, the realized standardized noise has mean $+0.146$ and standard deviation 0.671, below the law's unit standard deviation.
Both Gaussian forms place the noise scale at a posterior mean of \qty{0.271}{\mm}, close to $0.671\times\qty{0.4}{\mm}$.
\Cref{tab:sm_skew} gives $W_1/\mathrm{sd}$ for every parameter.
Outside the noise scale, the learned component stays below the split-half level of 0.125 in every parameter, and the Gaussian special case exceeds it in $\alpha_1$, $z_{l_c,1}$ and $z_{l_c,2}$.

\begin{table}[tbhp]
 \centering
 \caption{Skewed noise at the validation design: $W_1/\mathrm{sd}$ against direct sampling with the exact skewed likelihood.
 Learned: the component of \cref{sec:component} trained on the re-noised forward solves.
 Gaussian: the two forms of \cref{tab:sm_w1sd_valid} fitted to the same solves.
 Split-half level: the largest $W_1/\mathrm{sd}$ between the two halves of the direct sample.}
 \label{tab:sm_skew}
 \begin{tabular}{@{}lrrr@{}}
\toprule
 & & \multicolumn{2}{c}{Gaussian} \\ \cmidrule(lr){3-4}
parameter & learned & special case & mean refitted \\
\midrule
$\mu_{l_c}$ & 0.07 & 0.05 & 0.13 \\
$\sigma_{l_c}$ & 0.02 & 0.06 & 0.02 \\
$z_{l_c,1}$ & 0.04 & 0.13 & 0.04 \\
$z_{l_c,2}$ & 0.02 & 0.14 & 0.07 \\
$\alpha_{1}$ & 0.04 & 0.15 & 0.14 \\
$\alpha_{2}$ & 0.03 & 0.10 & 0.10 \\
$\sigma_{\Delta V}$ & 0.05 & 0.04 & 0.02 \\
$z_{\Delta V,1}$ & 0.02 & 0.02 & 0.04 \\
$z_{\Delta V,2}$ & 0.01 & 0.03 & 0.04 \\
$r_k$ & 0.08 & 0.07 & 0.14 \\
$z_k$ & 0.07 & 0.02 & 0.03 \\
$\Delta V_k$ & 0.10 & 0.05 & 0.05 \\
$\sigma_\text{exp}$ & 0.25 & 0.94 & 0.94 \\
\midrule
$\sigma_\text{exp}$, posterior mean [mm] & 0.350 & 0.271 & 0.271 \\
$\sigma_\text{exp}$, posterior mean, direct [mm] & \multicolumn{3}{c}{0.333} \\
split-half level of direct sampling, max & \multicolumn{3}{c}{0.125} \\
\bottomrule
\end{tabular}

\end{table}

\paragraph{Cost}
\Cref{tab:sm_cost} counts forward-model evaluations.
The second-to-last row prices a direct chain that stops once its slowest parameter reaches a bulk effective sample size of 200: the whole validation chain reaches \num{6134} in \num{250000} steps, and a linear extrapolation gives the length shown.
This rate includes the early steps of the chain.
At the rate of its last \num{95000} steps, a chain past its burn-in reaches 200 in about \num{5300} steps, and the ratio, with the burn-in left uncounted, falls to $10^{2.6}$.

\begin{table}[tbhp]
 \centering
 \caption{Forward-model evaluations of direct sampling against the training solves of the map, per posterior.
 Ratio: direct evaluations over training solves.
 Second-to-last row: a direct chain at the validation design that stops at a bulk effective sample size of 200.
 Last row: a calibration study of 200 such posteriors, in which both columns add one solve per cell for each simulated dataset; redrawn ground truths add solves for about \qty{0.2}{\percent} of the datasets at this design (\cref{tab:sm_domain}).}
 \label{tab:sm_cost}
 \begin{tabular}{@{}lrrrrrr@{}}
\toprule
 & & \multicolumn{3}{c}{direct sampling} & map & \\ \cmidrule(lr){3-5}\cmidrule(lr){6-6}
design & cells & steps & walkers & evaluations & solves & ratio \\
\midrule
validation & 4 & \num{250000} & 40 & $4.00\times10^{7}$ & \num{1990} & $10^{4.30}$ \\
scaling & 18 & \num{250000} & 40 & $1.80\times10^{8}$ & \num{1999} & $10^{4.95}$ \\
real data & 6 & \num{120000} & 40 & $2.88\times10^{7}$ & \num{1933} & $10^{4.17}$ \\
\midrule
validation, ESS 200 & 4 & \num{8151} & 40 & $1.30\times10^{6}$ & \num{1990} & $10^{2.82}$ \\
validation, 200 posteriors & 4 & $200\times\num{8151}$ & 40 & $2.61\times10^{8}$ & \num{2790} & $10^{4.97}$ \\
\bottomrule
\end{tabular}

\end{table}

\Cref{tab:sm_wallclock} gives the wall-clock time of the same comparison, as a secondary measure that also depends on the implementation and the hardware.

\begin{table}[tbhp]
 \centering
 \caption{Wall-clock time and process-hours of one posterior and of 200 posteriors at the validation design, both samplers stopped at a bulk effective sample size of 200.
 Direct sampling: the ensemble sampler with 40 walkers on 20 processes, at the time per step of the last \num{95000} steps of the validation chain, with the chain length of the second-to-last row of \cref{tab:sm_cost}.
 Map: the forward solves of the training rows and the training in one process, counted once.
 The No-U-Turn sampler of the map runs four chains on four processes, with its warm-up kept and its sampling shortened in proportion to its smallest bulk effective sample size.
 All runs on one 96-core server.
 Ratio: direct sampling over the map.
 The columns of 200 posteriors scale one posterior of the dataset of \cref{sec:synthetic}.
 In the validation study of \cref{sec:sbcresults}, the 196 replications that finish took 0.4 to 15 hours of wall-clock time each, with a median of 1.6 hours; each ran its four chains in one process, with 88 replications at a time on the same server.}
 \label{tab:sm_wallclock}
 \begin{tabular}{@{}lrrrr@{}}
\toprule
 & \multicolumn{2}{c}{wall-clock (h)} & \multicolumn{2}{c}{process-hours} \\ \cmidrule(lr){2-3}\cmidrule(lr){4-5}
& one & 200 & one & 200 \\
\midrule
direct sampling & \num{2.34} & \num{468} & \num{46.8} & \num{9360} \\
map & \num{1.08} & \num{26.2} & \num{1.46} & \num{102} \\
\midrule
ratio & \num{2.17} & \num{17.8} & \num{32.1} & \num{91.7} \\
\bottomrule
\end{tabular}

\end{table}

\begin{figure}[tbhp]
 \centering
 \includegraphics[width=\textwidth]{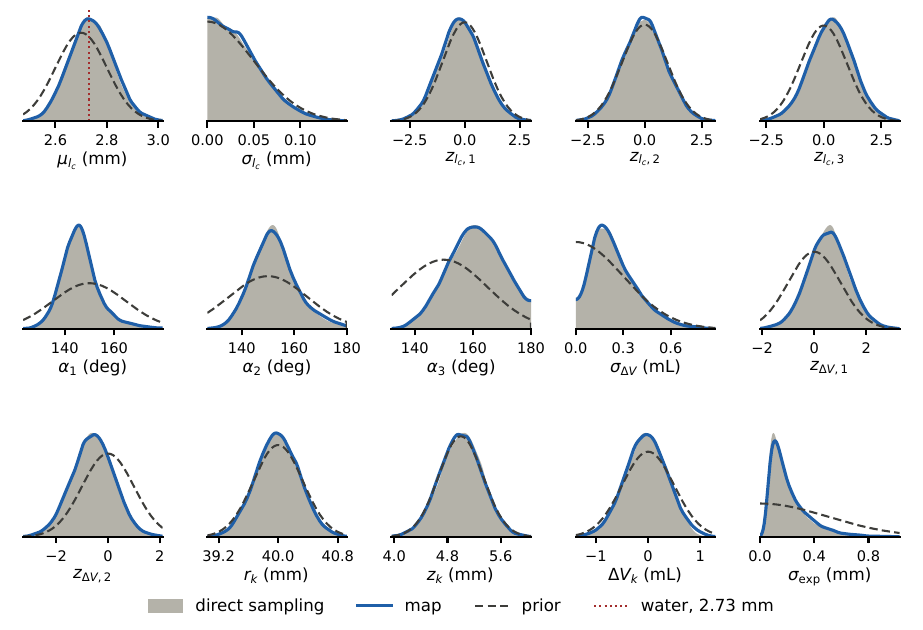}
 \caption{Posterior marginals at the real-data design, one panel per parameter.
 Shaded: direct sampling.
 Solid line: the map.
 Dashed line: the prior.
 Dotted line: the capillary length of water at \qty{20}{\celsius}.}
 \label{fig:sm_realdata}
\end{figure}

\paragraph{Real-data posterior}
\Cref{fig:sm_realdata} shows every marginal at the real-data design.
\Cref{tab:sm_real} gives $W_1/\mathrm{sd}$ for the four maps of \cref{sec:realposterior}, the split-half level of direct sampling, and the ratio of the map's posterior standard deviation to the prior standard deviation.
For a contact angle, the prior standard deviation is that of the truncated normal of \cref{tab:priors}.
For a capillary length $l_{c,i}$, it is $\sqrt{(\qty{0.1}{\mm})^2+(\qty{0.05}{\mm})^2}$, from the prior on $\mu_{l_c}$ and the half-normal prior on $\sigma_{l_c}$.

\begin{table}[tbhp]
 \centering
 \caption{The real-data design.
 $W_1/\mathrm{sd}$ against direct sampling for the map and its three replicate maps (rep.~1 to 3), the split-half level of direct sampling, and the posterior standard deviation under the map over the prior standard deviation.}
 \label{tab:sm_real}
 \begin{tabular}{@{}lrrrrrr@{}}
\toprule
 & \multicolumn{4}{c}{$W_1/\mathrm{sd}$, learned component} & direct & map \\ \cmidrule(lr){2-5}\cmidrule(lr){6-6}\cmidrule(lr){7-7}
parameter & map & rep.~1 & rep.~2 & rep.~3 & split half & sd ratio \\
\midrule
$\mu_{l_c}$ & 0.05 & 0.05 & 0.02 & 0.08 & 0.030 & -- \\
$\sigma_{l_c}$ & 0.01 & 0.01 & 0.02 & 0.02 & 0.070 & -- \\
$z_{l_c,1}$ & 0.02 & 0.03 & 0.01 & 0.02 & 0.042 & -- \\
$z_{l_c,2}$ & 0.03 & 0.02 & 0.03 & 0.02 & 0.028 & -- \\
$z_{l_c,3}$ & 0.02 & 0.01 & 0.01 & 0.01 & 0.063 & -- \\
$\alpha_{1}$ & 0.02 & 0.02 & 0.02 & 0.04 & 0.066 & 0.56 \\
$\alpha_{2}$ & 0.02 & 0.03 & 0.02 & 0.03 & 0.035 & 0.62 \\
$\alpha_{3}$ & 0.01 & 0.02 & 0.02 & 0.03 & 0.076 & 0.66 \\
$\sigma_{\Delta V}$ & 0.02 & 0.03 & 0.05 & 0.05 & 0.066 & -- \\
$z_{\Delta V,1}$ & 0.02 & 0.02 & 0.02 & 0.05 & 0.076 & -- \\
$z_{\Delta V,2}$ & 0.01 & 0.01 & 0.02 & 0.01 & 0.032 & -- \\
$r_k$ & 0.01 & 0.03 & 0.01 & 0.01 & 0.012 & -- \\
$z_k$ & 0.04 & 0.02 & 0.02 & 0.03 & 0.011 & -- \\
$\Delta V_k$ & 0.03 & 0.05 & 0.04 & 0.05 & 0.075 & -- \\
$\sigma_\text{exp}$ & 0.08 & 0.13 & 0.04 & 0.04 & 0.050 & -- \\
\midrule
$l_{c,1}$ & \multicolumn{5}{c}{} & 0.86 \\
$l_{c,2}$ & \multicolumn{5}{c}{} & 0.84 \\
$l_{c,3}$ & \multicolumn{5}{c}{} & 0.83 \\
\bottomrule
\end{tabular}

\end{table}

\FloatBarrier

\section{Calibration statistics}\label{sm:sbc}
\begin{figure}[tbhp]
 \centering
 \includegraphics[width=\textwidth]{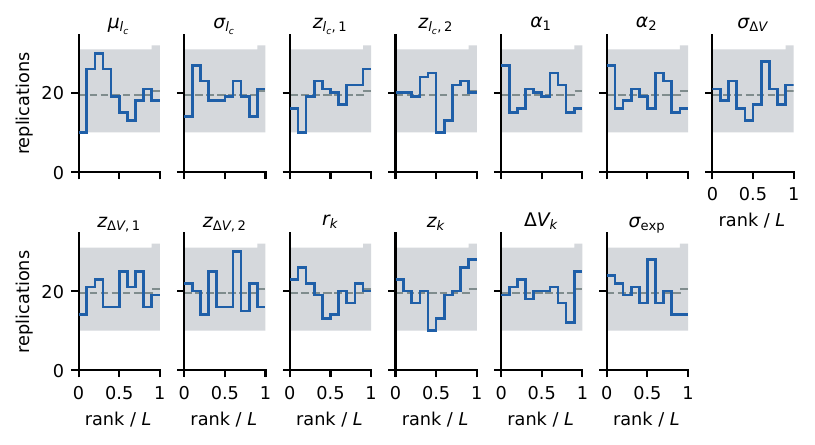}
 \caption{Rank histograms at the validation design, one panel per parameter of \cref{tab:priors}, drawn as in \cref{fig:sbcranks}.
 The band spans 10 to 31 counts, and 10 to 32 in the last bin.
 Four bins lie at its lower limit, one each of $\mu_{l_c}$, $z_{l_c,1}$, $z_{l_c,2}$ and $z_k$.}
 \label{fig:sm_sbcranks}
\end{figure}

\Cref{tab:sm_sbc} gives, for each parameter, the $p$-values of the chi-squared and Kolmogorov--Smirnov tests of \cref{sec:sbc} on the replications that pass the convergence gate, and \cref{tab:sm_sbc_ecdf} those of the ECDF test.
On all replications that finish, without the gate, every study has the same departures.
\Cref{fig:sm_sbcranks} shows the rank histograms of every parameter at the validation design.

\begin{table}[tbhp]
 \centering
 \caption{Simulation-based calibration, gated replications: $p$-values of the chi-squared test on ten bins and of the Kolmogorov--Smirnov test.
 Real data: two independent studies.
 A star marks a departure, a $p$-value below $0.05/n$.}
 \label{tab:sm_sbc}
 \begin{tabular}{@{}lrrrrrr@{}}
\toprule
 & \multicolumn{2}{c}{validation} & \multicolumn{2}{c}{real, study 1} & \multicolumn{2}{c}{real, study 2} \\ \cmidrule(lr){2-3}\cmidrule(lr){4-5}\cmidrule(lr){6-7}
parameter & $\chi^2$ $p$ & KS $p$ & $\chi^2$ $p$ & KS $p$ & $\chi^2$ $p$ & KS $p$ \\
\midrule
$\mu_{l_c}$ & 0.03 & 0.15 & 0.15 & 0.30 & 0.07 & $4.0\times10^{-3}$ \\
$\sigma_{l_c}$ & 0.58 & 0.78 & 0.71 & 0.20 & 0.36 & 0.22 \\
$z_{l_c,1}$ & 0.49 & 0.16 & 0.03 & 0.14 & 0.52 & 0.67 \\
$z_{l_c,2}$ & 0.32 & 0.55 & 0.35 & 0.52 & 0.13 & 0.55 \\
$z_{l_c,3}$ & -- & -- & 0.18 & 0.85 & 0.92 & 0.74 \\
$\alpha_{1}$ & 0.48 & 0.43 & 0.28 & 0.17 & 0.63 & 0.98 \\
$\alpha_{2}$ & 0.48 & 0.65 & 0.45 & 0.81 & 0.83 & 0.40 \\
$\alpha_{3}$ & -- & -- & 0.58 & 0.63 & 0.05 & 0.62 \\
$\sigma_{\Delta V}$ & 0.51 & 0.58 & 0.96 & 0.87 & 0.04 & 0.08 \\
$z_{\Delta V,1}$ & 0.58 & 0.77 & 0.30 & 0.12 & 0.06 & 0.45 \\
$z_{\Delta V,2}$ & 0.18 & 0.97 & 0.85 & 0.99 & 0.24 & 0.68 \\
$r_k$ & 0.58 & 0.11 & $1.4\times10^{-3}$$^\ast$ & 0.07 & 0.98 & 0.70 \\
$z_k$ & 0.18 & 0.07 & 0.16 & 0.05 & 0.65 & 0.12 \\
$\Delta V_k$ & 0.81 & 0.97 & 0.08 & 0.04 & $4.1\times10^{-3}$ & $7.2\times10^{-4}$$^\ast$ \\
$\sigma_\text{exp}$ & 0.40 & 0.17 & 0.10 & 0.03 & 0.55 & 0.64 \\
\midrule
replications, gated & \multicolumn{2}{c}{196} & \multicolumn{2}{c}{195} & \multicolumn{2}{c}{188} \\
replications, finished & \multicolumn{2}{c}{196} & \multicolumn{2}{c}{200} & \multicolumn{2}{c}{200} \\
threshold $0.05/n$ & \multicolumn{2}{c}{0.0038} & \multicolumn{2}{c}{0.0033} & \multicolumn{2}{c}{0.0033} \\
\bottomrule
\end{tabular}

\end{table}

\begin{table}[tbhp]
 \centering
 \caption{The ECDF test of S\"ailynoja et al.~\cite{Sailynoja2022-gu} on the replications of \cref{tab:sm_sbc}: Monte Carlo $p$-value of its statistic against \num{200000} uniform rank samples of the same size.
 A star marks a departure at the level $0.05/n$ of \cref{tab:sm_sbc}.}
 \label{tab:sm_sbc_ecdf}
 \begin{tabular}{@{}lrrr@{}}
\toprule
parameter & validation & real, study 1 & real, study 2 \\
\midrule
$\mu_{l_c}$ & $8.7\times10^{-3}$ & 0.084 & 0.010 \\
$\sigma_{l_c}$ & 0.89 & 0.46 & 0.23 \\
$z_{l_c,1}$ & 0.18 & 0.16 & 0.82 \\
$z_{l_c,2}$ & 0.88 & 0.73 & 0.41 \\
$z_{l_c,3}$ & -- & 0.59 & 0.82 \\
$\alpha_{1}$ & 0.18 & 0.30 & 0.97 \\
$\alpha_{2}$ & 0.29 & 0.91 & 0.50 \\
$\alpha_{3}$ & -- & 0.84 & 0.38 \\
$\sigma_{\Delta V}$ & 0.39 & 0.71 & 0.11 \\
$z_{\Delta V,1}$ & 0.67 & 0.23 & 0.23 \\
$z_{\Delta V,2}$ & 0.56 & 0.83 & 0.57 \\
$r_k$ & 0.30 & 0.051 & 0.61 \\
$z_k$ & 0.094 & 0.14 & 0.21 \\
$\Delta V_k$ & 1.00 & 0.015 & $2.0\times10^{-3}$$^\ast$ \\
$\sigma_\text{exp}$ & 0.17 & 0.098 & 0.54 \\
\bottomrule
\end{tabular}

\end{table}

At the validation design, with the ranks of the four replications that do not finish within \qty{16}{\hour} placed at 0 or at $L$, the smallest chi-squared or Kolmogorov--Smirnov $p$-value over all parameters, gated or ungated, is 0.023, above the threshold $0.05/13$.
At the uncorrected level 0.05, the ECDF test also rejects $\mu_{l_c}$ at the validation design ($p=0.0087$) and in real-data study 2, and $\Delta V_k$ in study 1.

\Cref{tab:sm_tertiles} gives the split of the validation study by the true noise scale (\cref{sec:sbcresults}).

\begin{table}[tbhp]
 \centering
 \caption{The validation study split into thirds by the true noise scale, finished replications: range of the true $\sigma_\text{exp}$, the mean of $z$, the true value minus the posterior mean over the posterior standard deviation, and the Kolmogorov--Smirnov $p$-value of the noise scale's ranks within each third.}
 \label{tab:sm_tertiles}
 \begin{tabular}{@{}lrrrr@{}}
\toprule
third & true $\sigma_\text{exp}$ [mm] & replications & mean $z$ & KS $p$ \\
\midrule
lower & 0.01--0.22 & 66 & $+0.06$ & 0.528 \\
middle & 0.22--0.43 & 65 & $-0.41$ & 0.011 \\
upper & 0.43--1.28 & 65 & $-0.09$ & 0.493 \\
\bottomrule
\end{tabular}

\end{table}

\paragraph{The departure in $\Delta V_k$}
\Cref{tab:sm_domain} gives the prior mass outside the domain of \cref{sec:rom} for \num{1500} prior draws per design.
At the real-data design the mass is largest, and the draws outside the domain have a larger $\Delta V_k$ than those inside.
\Cref{tab:sm_dvk} gives per study $z$, the true $\Delta V_k$ minus its posterior mean, over the posterior standard deviation, and the coverage of the central \qty{90}{\percent} interval.
\Cref{tab:sm_dvk} also gives the coverage of the central \qty{90}{\percent} credible interval of the noise scale, the share of truths strictly inside it.
For $\mu_{l_c}$, the mean of the same $z$ is $-0.02$, $-0.08$ and $-0.17$ in the validation study and real-data studies 1 and 2, each with a standard error of about 0.07, so only in real-data study 2 does its posterior mean lie clearly above the truth on average.

\begin{table}[tbhp]
 \centering
 \caption{Prior mass outside the domain, from \num{1500} prior draws per design, and the mean $\Delta V_k$ of the draws inside and outside the domain (prior mass only at the scaling design).}
 \label{tab:sm_domain}
 \begin{tabular}{@{}lrrr@{}}
\toprule
 & & \multicolumn{2}{c}{mean $\Delta V_k$ [mL]} \\ \cmidrule(lr){3-4}
design & outside [\%] & inside & outside \\
\midrule
validation & 0.2 & $+0.01$ & $+0.22$ \\
scaling & 0.7 & -- & -- \\
real data & 10.5 & $-0.08$ & $+0.62$ \\
\bottomrule
\end{tabular}

\end{table}

\begin{table}[tbhp]
 \centering
 \caption{Coverage of the central \qty{90}{\percent} interval for the noise scale, and for $\Delta V_k$ with the mean and standard deviation of $z$, in each calibration study (gated replications).}
 \label{tab:sm_dvk}
 \begin{tabular}{@{}lrrrr@{}}
\toprule
 & $\sigma_\text{exp}$ & \multicolumn{3}{c}{$\Delta V_k$} \\ \cmidrule(lr){2-2}\cmidrule(lr){3-5}
study & coverage & mean $z$ & sd $z$ & coverage \\
\midrule
validation & 0.888 & $-0.018$ & 1.009 & 0.893 \\
real data, study 1 & 0.887 & $-0.125$ & 0.862 & 0.918 \\
real data, study 2 & 0.888 & $-0.248$ & 0.835 & 0.963 \\
\bottomrule
\end{tabular}

\end{table}

\FloatBarrier

\section{The real-data cells}\label{sm:cells}
\Cref{tab:sm_cells} lists the six cells used.
The nominal volume $V_{0j}$ of a session is its mean dispensed volume over substrates minus the volume $V_k=\qty{4.835}{\ml}$ below the knife edge.
The dispensed volumes of one session differ from its mean by at most about \qty{0.05}{\ml}, small against the scale \qty{0.3}{\ml} of the prior on $\sigma_{\Delta V}$.

\begin{table}[tbhp]
 \centering
 \caption{The six cells used, one photograph each, with the nominal volume of its session. Substrates are numbered by their day of measurement, as in \cref{fig:experimentaldata}(a).}
 \label{tab:sm_cells}
 \begin{tabular}{@{}rlrrr@{}}
\toprule
substrate & day & session & dispensed [mL] & $V_{0j}$ [mL] \\
\midrule
1 & 230509 & 1 & 21.99 & 17.18 \\
2 & 230512 & 1 & 22.07 & 17.18 \\
3 & 230519 & 1 & 21.99 & 17.18 \\
1 & 230509 & 2 & 24.04 & 19.21 \\
2 & 230512 & 2 & 24.07 & 19.21 \\
3 & 230519 & 2 & 24.02 & 19.21 \\
\bottomrule
\end{tabular}

\end{table}

\FloatBarrier

\section{The standard component}\label{sm:stdcomp}
The standard component of \cref{sec:component}, of total order three with the SoftPlus rectifier and the noise scale itself as an input, was fitted to the same forward solves as the maps of \cref{sec:training}, with one noisy radius per solve and the noise scale drawn from its prior.
With the noise scale itself as an input, the polynomial $\partial_{y_n}f$ stays bounded as $\sigma_\text{exp}\to0$ while its target in \cref{sec:component} diverges, so the argument there holds as well.
Against direct sampling, its posterior places the noise scale low at the validation and scaling designs ($W_1/\mathrm{sd}=1.95$ and $6.13$) and high at the real-data design ($0.77$), and every other parameter stays within $0.08$, $0.21$ and $0.27$.
In simulation-based calibration studies of 300 replications, gated as in \cref{sec:sbc}, the noise scale departs at the validation and real-data designs (chi-squared 65.59 and 36.60 on nine degrees of freedom, against Bonferroni thresholds of 24.30 and 24.69), and the population mean of the capillary length at the real-data design (28.24).
The training data differ as well as the component, so the comparison shows that this standard component fails on this problem, and \cref{sec:component} gives the reason.

\end{document}